\documentclass[numberedappendix]{emulateapj}
\usepackage{epsfig}
\usepackage{amsfonts}
\usepackage{amsmath}
\usepackage{mathrsfs}
\usepackage{amssymb}
\usepackage{psfrag}
\bibpunct{(}{)}{;}{a}{}{,}

\def \kms            {{\rm km~s}^{-1}}
\def \skm            {{\rm s~km}^{-1}}

\newcommand{\mlya}{${\rm Ly\alpha}$}
\def\be{\begin{equation}}
\def\ee{\end{equation}}
\def\bea{\begin{eqnarray}}
\def\eea{\end{eqnarray}}

\submitted{}

\begin{document}
\title{A New Method to Directly Measure the Jeans Scale of the Intergalactic Medium Using Close Quasar Pairs}
\author{Alberto Rorai\altaffilmark{1,2}, Joseph F. Hennawi\altaffilmark{1} , Martin White\altaffilmark{3}}
\altaffiltext{1}{Max-Planck-Institut f\"ur Astronomie, K\"onigstuhl}
\altaffiltext{2}{International Max Planck Research School for Astronomy \& Cosmic Physics at the University of Heidelberg}
\altaffiltext{3}{Department of Astronomy, University of California at
  Berkeley, 601 Campbell Hall, Berkeley, CA 94720-3411}
\begin{abstract}

Although the baryons in the intergalactic medium (IGM) trace dark
matter fluctuations on Mpc scales, on smaller scales $\sim 100\,{\rm
  kpc}$, fluctuations are suppressed because the finite temperature gas is 
pressure supported against gravity,  analogous to the classical Jeans argument. 
This Jeans filtering scale, which quantifies the small-scale structure of the
IGM, has fundamental cosmological implications.  First, it provides a
thermal record of heat injected by ultraviolet photons during cosmic
reionization events, and thus constrains the thermal and reionization
history of the Universe. Second, the Jeans scale determines the
clumpiness of the IGM, a critical ingredient in models of cosmic
reionization. Third, it sets the minimum mass scale for gravitational
collapse from the IGM, and hence plays a pivotal role in galaxy
formation. Unfortunately, it is extremely challenging to measure the
Jeans scale via the standard technique of analyzing purely
longitudinal \mlya\ forest spectra, because the thermal Doppler
broadening of absorption lines along the line-of-sight, is highly degenerate with Jeans
smoothing. In this work we show that the Jeans filtering scale can be
directly measured by characterizing the coherence of correlated
\mlya\ forest absorption in close quasar pairs, with separations small
enough $\sim 100\,$kpc to resolve it. We present a novel technique for
this purpose, based on the probability distribution function (PDF) of
phase angle differences of homologous longitudinal Fourier modes in
close quasar pair spectra. A Bayesian formalism is introduced based on 
the phase angle PDF, and MCMC techniques are used to characterize the
precision of a hypothetical Jeans scale measurement, and explore
degeneracies with other thermal parameters governing the IGM.  A
semi-analytical model of the \mlya\ forest is used to generate a large
grid ($500$) of thermal models from a dark matter only simulation.
Our full parameter study indicates that a realistic sample of only 20
close quasar pair spectra can pinpoint the Jeans scale to $\simeq 5\%$
precision, independent of the amplitude $T_0$ and slope $\gamma$
of the temperature-density relation of the IGM $T=T_0(\rho\slash {\bar
  \rho})^{\gamma-1}$. This exquisite sensitivity arises because even
long-wavelength 1D Fourier modes $\sim 10\,$Mpc, i.e.  two orders of
magnitude larger than the Jeans scale, are nevertheless dominated by
projected small-scale 3D power. Hence phase angle differences
between \emph{all} modes of quasar pair spectra actually probe the
shape of the 3D power spectrum on scales comparable to the pair
separation. We show that this new method for measuring the Jeans scale
is unbiased and is insensitive to a battery of systematics that typically
plague Ly$\alpha$ forest measurements, such as continuum fitting
errors, imprecise knowledge of the noise level and/or spectral
resolution, and metal-line absorption.

\end{abstract}
\section{Introduction}
The imprint of redshifted Lyman-$\alpha$ (Ly$\alpha$) forest
absorption on the spectra of distant quasars provides an exquisitely
sensitive probe of the distribution of baryons in the intergalactic
medium (IGM) at large cosmological lookback times. Among the
remarkable achievements of modern cosmology is the ability of
cosmological hydrodynamical simulations to explain the origin of this
absorption pattern, and reproduce its statistical properties to
percent level accuracy \citep[e.g.][]{Cen94,MEscude96,Rauch98}. But
the wealth of information which can be gathered from the Ly$\alpha$
forest is far from being exhausted.  The thermal state of the baryons
in the IGM reflects the integrated energy balance of heating --- due
to the collapse of cosmic structures, radiation, and possibly other
exotic heat sources --- and cooling due to the expansion of the
Universe \citep[e.g.][]{miraldarees94,HuiGnedin97,HH03,Meiksin09}.  Cosmologists
still do not understand how the interplay of these physical processes
sets the thermal state of the IGM, nor has this thermal state been 
precisely measured.

There is ample observational evidence that ultraviolet radiation
emitted by the first star-forming galaxies ended the `cosmic dark
ages' ionizing hydrogen and singly ionizing helium at $z\sim 10$
\citep[e.g.][]{BarkanaLoeb01,Ciardi05,Fan06,Zaroubi2013}.  A second
and analogous reionization episode is believed to have occurred at
later times $z\sim 3-4$ \citep{MadauMeiksin94,Jacobsen94,Reimers97,Croft97},
when quasars were sufficiently abundant to supply the hard photons
necessary to doubly ionized helium. The most recent observations from
HST/COS provide tentative evidence for an extended \ion{He}{2}
reionization from $z\sim 2.7-4$
\citep[][Worseck et al. 2013, in preparation]{Shull10,Furl10,Worseck2011}, with a duration of
$\sim 1$\,Gyr, longer than naively expected. Cosmic reionization
events are watersheds in the thermal history of the Universe,
photoheating the IGM to tens of thousands of degrees. Because cooling
times in the rarefied IGM gas are long, memory of this heating is
retained
\citep{miraldarees94,HuiGnedin97,Haehnelt98,HH03,Theuns02a,Theuns02b}.
Thus an empirical characterization of the IGMs thermal history
constrains the nature and timing of reionization.

From a theoretical perspective, the impact of reionization events on
the thermal state of the IGM is poorly understood.  Radiative transfer
simulations of both hydrogen \citep{Bolton04,Iliev06,Titley07} and
helium \citep{Abel99,McQuinn09,MeiksinTittley12} reveal that the heat
injection and the resulting temperature evolution of the IGM depends
on the details of how and when reionization occurred. There is
evidence that the thermal vestiges of \ion{H}{1} reionization heating
may persist until as late as $z\sim 4-5$, and thus be observable in
the Ly$\alpha$ forest \citep{HH03,FurlanettoOh09,Cen09}, whereas for
HeII reionization at $z\sim 3$, the Ly$\alpha$ forest is observable
over the full duration of the phase transition.  Finally, other
processes could inject heat into the IGM and impact its thermal
state, such as the large-scale structure shocks which eventually
produce the Warm Hot Intergalactic Medium
\citep[WHIM;e.g.][]{CenOstriker1999,Dave99,Dave01}, heating from
galactic outflows \citep{Kollmeier06,CenOstriker06}, photoelectric
heating of dust grains \citep{Nath99,Inoue03}, cosmic-ray heating
\citep{Nath93}, Compton-heating from the hard X-ray background
\citep{MadauEf99}, X-ray preheating \citep{ROG05,Tanaka12}, or blazar
heating \citep{Blazar1,Blazar2,Blazar3,Puchwein12}.  Precise
constraints on the thermal state of the IGM would help determine the
relative importance of photoheating from reionization and these more
exotic mechanisms.

Despite all the successes of our current model of the IGM, precise
constraints on its thermal state and concomitant constraints on
reionization (and other exotic heat sources) remain elusive. Attempts
to characterize the IGM thermal state from Ly$\alpha$ forest
measurements have a long history.  In the simplest picture, the gas in
the IGM obeys a power law temperature-density relation $T =
T_0(\rho\slash {\bar \rho})^{\gamma-1}$, which arises from the balance
between photoionization heating, and cooling due to adiabatic
expansion \citep{HuiGnedin97}. The standard approach has been to
compare measurements of various statistics of the Ly$\alpha$ forest to
cosmological hydrodynamical simulations. Leveraging the dependence of
these statistics on the underlying temperature-density relation, its
slope and amplitude $(T_0,\gamma)$ parameters can be constrained.  To
this end a wide variety of statistics have been employed, such as the
power spectrum \citep{Zald01,VielBolton09} or analogous statistics quantifying the
small-scale power like wavelets \citep{Theuns02b,Lidz09,Garzilli2012}
or the curvature \citep{BeckerBolton2011}.  The flux PDF
\citep{McDonald2000,kbv+07,Bolton08,Calura2012,Garzilli2012} and the shape of the
$b$-parameter distribution
\citep{Haehnelt98,Theuns00,Ricotti00,BryanMach00,Schaye00,McDonald2001,Theuns02a,Rudie2012}
have also been considered.  Multiple statistics have also been
combined such as the PDF and wavelets \citep{Garzilli2012}, or PDF and
power spectrum \citep{VielBolton09}.  Overall, the results of such
comparisons are rather puzzling. First, the IGM appears to be
generally too hot, both at low ($z \sim 2$) and high ($z\sim 4$)
redshift \citep{HH03}. In particular, the high inferred temperatures
at $z\sim 4$
\citep[e.g.][]{Schaye00,Zald01,McDonald2001,Theuns02b,Lidz09} suggest
that HeII was reionized at still higher redshift $z > 4$ \citep{HH03},
possibly conflicting with the late $z\sim 2.7$ reionization of HeII
observed in HST/COS spectra
\citep[][Worseck et al. 2013, in preparation]{Furl10,Shull10,Worseck2011,Syphers12}. Second, 
\citet{Bolton08} considered the PDF of high-resolution quasar
spectra and concluded that, at $z\simeq 3$ the slope of the
temperature-density relation $\gamma$ is either close to isothermal
($\gamma = 1$) or even inverted ($\gamma < 1$), suggesting ``that the
voids in the IGM may be significantly hotter and the thermal state of
the low-density IGM may be substantially more complex than is usually
assumed.''  Although this result is corroborated by additional work
employing different statistics/methodologies \citep[][but see Lee et
  al. 2012]{VielBolton09,Calura2012,Garzilli2012}, radiative
transfer simulations of HeII reionization cannot produce an isothermal
or inverted slope, unless a population other than quasars reionized
HeII \citep{Bolton04,McQuinn09,MeiksinTittley12} , which would fly in
the face of conventional wisdom. To summarize, despite nearly a decade
of theoretical and observational work, published measurements of the
thermal state of the IGM are still highly confusing, and concomitant
constraints on reionization scenarios are thus hardly compelling.

Fortunately, there is another important record of the thermal history
of the Universe: the Jeans pressure smoothing scale.  Although baryons
in the IGM trace dark matter fluctuations on large Mpc scales, on
smaller scales $\lesssim 100~{\rm kpc}$, gas is pressure
supported against gravitational collapse by its finite temperature.
Analogous to the classic Jeans argument, baryonic fluctuations are
suppressed relative to the pressureless dark matter (which can
collapse), and thus small-scale power is `filtered' from the IGM
\citep{GnedHui98}, which explains why it is sometimes referred to as the
\emph{filtering scale}. Classically the \emph{comoving} Jeans scale is
defined as $\lambda_J^{0}=\sqrt{\pi c_s^2 / G\rho}(1+z)$, but in reality
the amount of Jeans filtering is sensitive
to both the instantaneous pressure and hence temperature of the IGM,
\emph{as well as the temperature of the IGM in the past}. This arises because
fluctuations at earlier times expanded or failed to collapse depending
on the IGM temperature at that epoch. Thus the Jeans scale
reflects the competition between gravity and pressure integrated 
over the Universe's history,  and cannot be expressed as a mere deterministic function of the
instantaneous thermal state. Heuristically, this can be understood because reionization
heating is expected to occur on the reionization timescales of several 
hundreds of Myr, whereas the baryons respond to this heating on the
sound-crossing timescale $\lambda_J^{0}\slash [c_s(1+z)] \sim
\left(G\rho\right)^{-1\slash 2}$, which at mean density is comparable to the
Hubble time $t_H$.  

\citet{GnedHui98} considered the behavior of the Jeans smoothing in
linear theory, and derived an analytical expression for the filtering
scale $\lambda_J$ as a function of thermal history
\begin{equation}
\begin{split}
\left.    \lambda_J^2(t)=\frac{1}{D_{+}(t)}\int_0^t dt' a^2(t')(\lambda_J^{0}(t'))^2 \times \right. \\
\left.(\ddot{D}_+(t')+2H(t')\dot{D}_+(t'))\int_{t'}^{t}\frac{dt''}{a^2(t'')}  ,   \right.
\end{split} 
\label{eqn:jeans}
\end{equation}
where $D_+(t)$ is the linear growth function at time $t$, $a(t)$ is the
scale factor, and $H(t)$ the Hubble expansion rate. 
Although this simple linear approximation provides intuition about the
Jeans scale and its evolution, Fourier modes with wavelength
comparable to the Jeans scale are already highly nonlinear at $z\sim
3$, and hence this simple linear pictures breaks down due
to nonlinear mode-mode coupling effects. 
Thus given that we do not know the thermal history of the Universe,
that we expect significant heat injection from HeII reionization at
$z\sim 3-4$ concurrent with the epoch at which we observe the IGM, and
that IGM modes comparable to the Jeans scale actually respond
non-linearly to this unknown heating, the true relationship between the Jeans
scale and the temperature-density relation at a given epoch should be regarded as
highly uncertain.

Besides providing a thermal record of the IGM, the small-scale
structure of baryons, as quantified by the Jeans scale, is a fundamental
ingredient in models of reionization and galaxy formation. A
critical quantity in models of cosmic reionization is the 
clumping factor of the IGM $C=\langle n_H^2\rangle/\bar{n}_H^2$
\citep[e.g.][]{MadauHR99,MEscudeHaehnelt00,PawlikSchaye2009,Haardt12,Emberson13,McQuinn11},
because it determines the average number of recombinations per atom,
or equivalently the total number of UV photons needed to keep the IGM
ionized. The clumping and the Jeans scale are directly
related. Specifically, \be C = 1 + \sigma^2_{\rm IGM} \equiv 1 +
\int d\ln k \,\frac{k^3 P_{\rm IGM}(k)}{2\pi^2}\label{eqn:clump}, \ee
where $\sigma^2_{\rm IGM}$ is the variance of the IGM density, and
$P_{\rm IGM}(k)$ is the 3D power spectrum of the
baryons in the IGM. Given the shape of $P_{\rm IGM}(k)$, the
integral above is dominated by contributions from small-scales
(high-$k$), and most important is the Jeans cutoff $\lambda_J$, which
determines the maximum $k$-mode $k_{\rm J}\sim 1\slash \lambda_{\rm
  J}$ contributing. The small-scale structure of the IGM strongly
influences the propagation of cosmological ionization fronts during
reionization \citep{Iliev05}. Furthermore, several numerical studies
have revealed that the hydrodynamic response of the baryons in the IGM
to impulsive reionization heating is significant
\citep[e.g.][]{Gnedin00,Haiman01,Kuhlen05,Ciardi07,PawlikSchaye2009},
indicating that a full treatment of the interplay between IGM
small-scale structure and reionization history probably requires
coupled radiative transfer hydrodynamical simulations.

Reionization heating also evaporates the baryons from low-mass halos
or prevents gas from collapsing in them altogether
\citep[e.g.][]{BarkanaLoeb99,Dijkstra04}, an effect typically modeled
via a critical mass, below which galaxies cannot form
\citep{Gnedin2000,Bullock00,Benson02a,Benson02b,Somerville02,Kulkarni11}.
\citet{Gnedin2000} used hydrodynamical simulations to show that this
scale is well approximated by the \emph{filtering mass}, which is the
mass-scale corresponding to the Jeans filtering length,
i.e. $M_F(z)=4\pi{\bar \rho}\lambda^3_J/3$ \citep[see
  also][]{Hoeft06,Okamoto08}. Finally, because the Jeans scale has
memory of the thermal events in the IGM (see eqn.~\ref{eqn:jeans}),
its value at later times can potentially constrain models of early IGM
preheating. In this scenario, heat is globally injected into the IGM
at high-redshift $z\sim 5-15$ from blast-waves produced by outflows
from proto-galaxies or miniquasars
\citep{Voit96,Madau2000,Madau01,CenBryan01,TheunsMoSchaye01,BensonMadau03,Scannapieco02,Scannapieco04}
X-ray radiation from early miniquasars \citep{Tanaka2012,Parsons13},
which sets an entropy floor in the IGM and the raises filtering mass
scale inhibiting the formation of early galaxies.

A rough estimate of the filtering scale at $z=3$ can be obtained from
eqn.~(\ref{eqn:jeans}) and the following simplified assumptions: the
temperature at $z=3$ is $T(z=3) \approx 15000$\,K as suggested by
measurements \citep[e.g.][]{Schaye00,Ricotti00,Zald01,Lidz09},
temperature evolves as $T \propto 1+z$, the typical overdensity probed
by the $z=3$ \mlya\ forest is $\delta \sim 2$
\citep{BeckerBolton2011}.  One then obtains $\lambda_J(z=3) \approx
340$ kpc (comoving), smaller than the classical or instantaneous Jeans
scale $\lambda_J^0$ by a factor of $\sim 3$.  
This distance maps to a velocity interval
$v_J=Ha\lambda_J\approx 26\,\kms$ along the line of sight due to
Hubble expansion. Thermal Doppler broadening gives rise to a cutoff
in the longitudinal power spectrum, which occurs at a comparable
velocity $v_{\rm th}\approx 11.3\,\kms$, for gas heated to the same
temperature. The similarity of the characteristic scale of 3D Jeans
pressure smoothing and the 1D thermal Doppler smoothing suggests that
disentangling the two effects will be challenging given purely
longitudinal observations of the \mlya\ forest, as confirmed by
\cite{Peeples09a}, who considered the relative impact of thermal
broadening and pressure smoothing on various statistics applied to
longitudinal Ly$\alpha$ forest spectra.  Previous work that has aimed
to measure thermal parameters such as $T_0$ and $\gamma$ from
Ly$\alpha$ forest spectra, have largely ignored the degeneracy of the
Jeans scale with these thermal parameters. The standard approach has
been to assume values of the Jeans scale from a hydrodynamical
simulation \citep[e.g.][]{Lidz09,VielBolton09,BeckerBolton2011}, which
as per the discussion above, is equivalent to assuming perfect
knowledge of the IGM thermal history. Because of the degeneracy with
the Jeans scale, it is thus likely that previous measurements of the
thermal parameters $T_0$ and $\gamma$ are significantly biased, and
their error bars significantly underestimated, if indeed Jeans scale
takes on values different from those assumed (but see Zaldarriaga et
al. 2001 who marginalized over the Jeans scale, and Becker et al. 2011
who also considered its impact).  We will investigate such
degeneracies in detail in this paper with respect to power-spectra,
and we consider degeneracies for a broader range of IGM statistics in
a future work (A.Rorai et al. 2013, in preparation).

The Jeans filtering scale can be directly measured using close quasar
pair sightlines which have comparable transverse separations
$r_{\perp} \lesssim 300\,$kpc (comoving; $\Delta\theta \lesssim
40\arcsec$ at $z=3$). The observable signature of Jeans smoothing is
increasingly coherent absorption between spectra at progressively smaller pair
separations resolving it \citep{Peeples09b}.  The idea of using pairs
to constrain the small scale structure of the IGM is not new. However,
all previous measurements have either focused on lensed quasars, which
probe extremely small transverse distances $r_{\perp} \sim 1\,$kpc
$\ll \lambda_J$
\citep[e.g.][]{Young81,McGill90,PIF98,Smette95,Rauch01} such that the
Ly$\alpha$ forest is essentially perfectly coherent, or real physical quasar pairs
with $r_{\perp} \sim 1$ Mpc $\gg \lambda_J$ \citep{DOdorico06} far too
large to place useful constraints on the Jeans scale. Observationally,
the breakthrough enabling a measurement of the Jeans scale is the
discovery of a large number of close quasar pairs
\citep{Hennawi04,BINARY,Myers08,HIZBIN} with $\sim 100\,$kpc
separations. By applying machine
learning techniques \citep{richards04,Bovy11,Bovy12} to the Sloan Digital Sky
Survey \citep[SDSS;][]{yaa+00} imaging, a sample of $\sim 300$ close
$r_{\perp} < 700\,$kpc quasar pairs at $1.6 < z \lesssim
4.3$\footnote{The lower redshift limit is corresponds to Ly$\alpha$
  forest absorption being above the atmospheric cutoff.} has been
uncovered \citep{Hennawi04,BINARY,HIZBIN}.

In this paper we introduce a new method which will enable the first
determination of the Jeans scale, and we estimate the precision with
which it can be measured from this close quasar pair dataset. We
explicitly consider degeneracies between the canonical thermal
parameters $T_0$ and $\gamma$, and the Jeans scale $\lambda_J$, which
have been heretofore largely ignored. To this end, we use an
approximate model of the Ly$\alpha$ forest based on dark matter only
simulations, allowing us to independently vary all thermal parameters
and simulate a large parameter space.  The structure of this paper is
as follows: we describe how we compute the Ly$\alpha$ forest flux
transmission from dark matter simulations, and our parametrization of
the thermal state of the IGM in section \S~\ref{sim_met}.  In
\S~\ref{ps_cps} we consider thermal parameter degeneracies which
result when only longitudinal observations are available, and we show
how the additional transverse information provided by quasar pairs can
break them. In \S~\ref{phase_ang} we introduce our new method to
quantify absorption coherence using the difference in phase between
homologous longitudinal Fourier modes of each member of a quasar
pair. We focus on the probability distribution function (PDF) of
these phase differences, and find that the shape of this phase PDF is
very sensitive to the Jeans smoothing.  A Bayesian likelihood
formalism that uses the phase angle PDF to determine the Jeans scale
is presented in \S~\ref{jeans_meas}. Our Bayesian method allows us to
combine the Jeans scale information with other Ly$\alpha$ forest
statistics such as the longitudinal power spectrum, and we conduct a
Markov Chain Monte Carlo (MCMC) analysis in this section to determine
the resulting precision on $T_0$, $\gamma$, and $\lambda_J$ expected
for realistic datasets, explore parameter degeneracies, and study the
impact of noise and systematic errors.  We conclude and summarize in
\S~\ref{summary}.

Throughout this paper we use the $\Lambda$CDM cosmological model with the
parameters $\Omega_m=0.28, \Omega_{\Lambda}=0.72, h=0.70, n=0.96, \sigma_8=0.82 $. 
All distances quoted are in comoving kpc.

\section{Simulation Method}\label{sim_met}
\subsection{Dark Matter Simulation}
Our model of the Ly$\alpha$ forest is based on a single snapshot of a
dark matter only simulation at $z=3$.  In this scheme, the dark matter
simulation provides the dark matter density and velocity field
\citep{Croft98,MeiksinWhite2001}, and the gas density and temperature are
computed using simple scaling relations motivated by the results of
full hydrodynamical simulations
\citep{HuiGnedin97,GnedHui98,GnedinBaker2003}. Our objective is then
to explore the sensitivity with which close quasar pairs can be used to constrain
the thermal parameters defining these scaling relations, and in particular the 
Jeans scale. To this end, we require a dense sampling of the thermal
parameter space, which is computationally feasible with our
semi-analytical method applied to a dark matter simulation snapshot,
whereas it would be extremely challenging to simulate such a dense
grid with full hydrodynamical simulations. We do not model the
redshift evolution of the IGM, nor do we consider the effect of
uncertainties on the cosmological parameters, as they are constrained
by various large-scale structure and CMB measurements to much higher
precision than the thermal parameters governing the IGM.

We used an updated version version of the TreePM code described in
\citet{TreePM} to evolve $1500^3$ equal mass ($3\times
10^{6}\,h^{-1}M_\odot$) particles in a periodic cube of side length
$L_{\rm box}=50\,h^{-1}$Mpc with a Plummer equivalent smoothing of
$1.2\,h^{-1}$kpc.  The initial conditions were generated by displacing
particles from a regular grid using second order Lagrangian
perturbation theory at $z=150$.  This TreePM code has been compared to
a number of other codes and has been shown to perform well for such simulations
\citep{Hei08}.  Recently the code has been modified to use a hybrid
MPI+OpenMP approach which is particularly efficient for modern
clusters.

\subsection{Description of the Intergalactic Medium}
The baryon density field is obtained by smoothing the dark matter
distribution; this smoothing mimics the effect of the Jeans pressure
smoothing.  For any given thermal model, we adopt a constant filtering
scale $\lambda_J$, rather than computing it as a function of the
temperature, and this value is allowed to vary as a free parameter
(see discussion below). The dark matter distribution is convolved
with a window function $W_{\rm IGM}$, which, in
Fourier space,  has the effect of quenching high-$k$ modes 
\begin{equation}
 \delta_{\rm IGM}(\vec{k})=W_{\rm IGM}(\vec{k},\lambda_J)\delta_{\rm DM}(\vec{k}) 
\end{equation}
For example a Gaussian kernel with $\sigma=\lambda_J$,
$W_{\rm IGM}(k)=\exp (-k^2\lambda_J^2/2)$, would truncates the 3D power spectrum at $k \sim 1/\lambda_J$.

Because we smooth the dark matter particle distribution in real-space,  it is more convenient to adopt a 
function with a  finite-support 
\begin{equation}
\delta_{\rm IGM}(x) \propto \sum_i m_i K(|x-x_i|,R_J)
\end{equation}
where $m_i$ and $x_i$ are the mass and position of the particle $i$, $K(r)$ is the kernel, and $R_J$ the smoothing parameter which sets the Jeans scale. We adopt the followoing cubic spline kernel
\begin{equation}
K(r,R_J)=\frac{8}{\pi R_J^3}
  \begin{cases}
      1-6\left(\frac{r}{R_J}\right)^2+6\left(\frac{r}{R_J}\right)^3 & \frac{r}{R_J} \leq \frac{1}{2} \\ 
      2\left(1-\frac{r}{R_J}\right)^3 & \frac{1}{2}<\frac{r}{R_J}\leq 1 \\ 
      0 & \frac{r}{R_J} >1
  \end{cases}.\label{eqn:kernel}
\end{equation}
In the central regions the shape of $K(r)$ very closely resembles a Gaussian
with $\sigma \sim R_J/3.25 $, and we will henceforth take this $R_J/3.25$ to be our definition
of $\lambda_J$, which we will alternatively refer to as the `Jeans
scale' or the `filtering scale'.  The analogous smoothing procedure is also 
applied to the particle velocities; however, note that the velocity field has
very little small-scale power, and so the velocity distribution is 
essentially unaffected by this pressure smoothing operation. As we discuss further in 
Appendix \ref{sec:appendixa}, the mean inter-particle separation of our simulation cube $\delta l= L_{\rm box}\slash N_{\rm p}^{1\slash 3}$ sets the minimum Jean smoothing
that we can resolve with our dark matter simulation, hence we can safely model values
of $\lambda_J > 50\,{\rm kpc}$.

At the densities typically probed by the Ly$\alpha$ forest, the IGM is
governed by relatively simple physics. Most of the gas has never been shock heated,
is optically thin to ionizing radiation, and can be considered to be
in ionization equilibrium with a uniform UV background. Under these
conditions, the competition between photoionization heating and
adiabatic expansion cooling gives rise to a tight relation between
temperature and density which is well approximated by a power law
\citep{HuiGnedin97},
\begin{equation}
T(\delta)=T_0 (1+\delta)^{\gamma-1} \label{eqn:rhoT}
\end{equation}
where $T_0$, the temperature at the mean density, and $\gamma$, 
the slope of the temperature-density relation, both depend on the thermal history of the gas. 
We thus follow the standard approach, and parametrize the thermal state of the IGM in this way. 
Typical values for $T_0$ are on the order of $10^4$ K, while $\gamma$ is expected to be around
unity, and asymptotically approach the value of $\gamma_{\infty}=1.6$, if there is no other
heat injection besides (optically thin) photoionzation heating.
Recent work suggests that an inverted temperature-density relation $\gamma < 1$ 
provides a better match to the flux probability distribution of the Ly$\alpha$ forest \citep{Bolton08}, but 
the robustness of this measurement has been debated \citep{Lee2012}.

The optical depth for Ly$\alpha$ absorption is proportional to the density of
neutral hydrogen $n_{HI}$, which, if the gas is highly ionized ($x_{HI}\ll 1$)
and in photoionization equilibrium, can be calculated as \citep{gp65}
\begin{equation}
n_{HI} = \alpha(T) n_{H}^2/ \Gamma 
\end{equation}
where $\Gamma$ is the photoionization rate due to a uniform metagalactic ultraviolet background (UVB), 
and $\alpha(T)$ is the recombination coefficient which scales as $ T^{-0.7}$ at typical IGM temperatures. 
These approximations result in a power law relation between Ly$\alpha$ optical depth and
overdensity often referred as the fluctuating Gunn-Petersonn approximation (FGPA)
$\tau\propto (1+\delta)^{2-0.7(\gamma-1)}$, which does not include the effect of peculiar motions and thermal
broadening. We compute the observed optical depth in redshift-space via the following
convolution of the real-space optical depth 
\begin{equation}
 \tau(v)=\int_{-\infty}^{\infty} \tau(x) \Phi(Hax+v_{p,\parallel}(x)-v, b(x))dx \label{eqn:tau},
\end{equation}
where $Hax$ is the real-space position in velocity units,
$v_{p,\parallel}(x)$ is the longitudinal component of the peculiar
velocity of the IGM at location $x$, and $\Phi$ is the normalized
Voigt profile (which we approximate with a Gaussian) characterized by
the thermal width $b=\sqrt{2K_BT/mc^2}$, where we compute the
temperature from the baryon density via the temperature-density
relation (see eqn.~\ref{eqn:rhoT}).  The observed flux transmission is 
then given by $F(v)=e^{-\tau(v)}$.

We apply the aforementioned recipe to $2\times 100^2$ lines-of-sight
(\emph{skewers}) running parallel to the box axes, to generate the
spectra of $100^2$ quasar pairs, and we repeat this procedure for 500
different choices of the parameter set $(T_0,\gamma,\lambda_J)$.  Half
of the spectra (the first member of each pair) are positioned on a
regular grid in the $y-z$ plane, in order to distribute them evenly in
space. Subsequently, a companion is assigned to each of them, and our
choice for the distribution of radial distances warrants further
discussion. Our goal is to statistically characterize the coherence of
pairs of spectra as a function of impact parameter, and near the Jeans
scale this coherence varies rapidly with pair separation. Hence
computing statistics in bins of transverse separation is undesirable,
because it can lead to subtle biases in our parameter determinations
if the bins are too broad. To circumvent these difficulties, we focus
our entire analysis on 30 linearly-spaced discrete pair separations
between $0$ and $714$ kpc. For each of the $100^2$ lines-of-sight on
the regular grid, a companion sightline is chosen at one of these
discrete radial separations, where the azimuthal angle is drawn from a
uniform distribution.

\begin{figure*}
  \centering \centerline{\epsfig{file=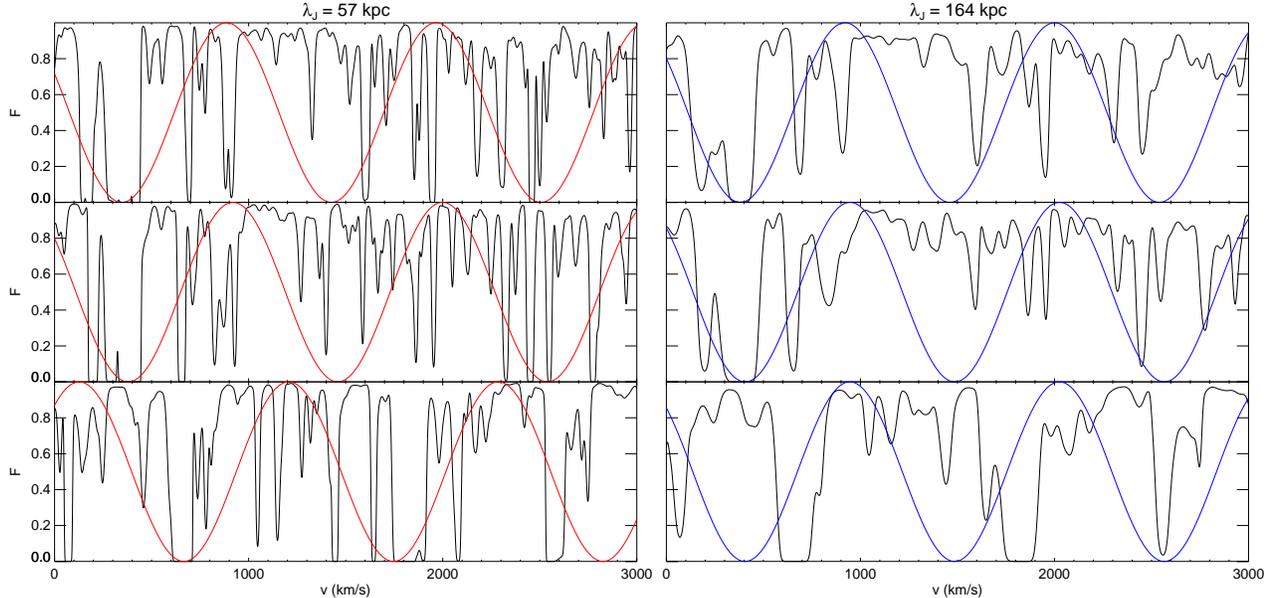,
      width=\textwidth}}
  \vskip -0.1in
  \caption{\label{fig:spec_w_phases} An example of three simulated
    spectra. The left and the right panels represent the same spectra
    in the simulation calculated for two models with different Jeans
    smoothing length $\lambda_J$.  The middle and the lower panel
    represent two spectra respectively at separation $0.5$ Mpc and $1$
    Mpc from the top one. The coloured sine curves track homologous
    Fourier modes in each spectrum, with rescaled mean and amplitude
    to fit the range $[0,1]$. The wave shifts provide a graphical
    visualization of phase differences, which we will use to quantify spectral 
    coherence and probe the Jeans scale. The right panels suggest that a larger 
    $\lambda_J$ results in greater spectral coherence and generally smaller
    phase differences between neighboring sightlines.}
\end{figure*}

We follow the standard approach, and treat the metagalactic
photoionization rate $\Gamma$ as a free parameter, whose value is
fixed \emph{a posteriori} by requiring the mean flux of our Ly$\alpha$
skewers $\langle \exp(-\tau)\rangle$ to match the measured values from
\cite{faucher08}.  This amounts to a simple constant re-scaling of the
optical depth.  The value of the mean flux at $z=3$ is taken to be
fixed, and thus assumed to be known with infinite precision. This is
justified, because in practice, the relative measurement errors on the
mean flux are very small in comparison to uncertainties of the
thermal parameters we wish to study. In a future work, we conduct a full parameter
study using other Ly$\alpha$ forest statistics, and explore the effect
of uncertainties of the mean flux (A.Rorai et al. 2013, in preparation).  
Examples of our spectra are shown in Figure~\ref{fig:spec_w_phases}.

To summarize, our models of the Ly$\alpha$ forest are uniquely described by the three parameters
($T_0, \gamma$, $\lambda_J$), and we reiterate that these three parameters
are considered to be independent.  In particular the Jeans scale
is not related to the instantaneous temperature at mean density $T_0$. Although this may at 
first appear unphysical, it is motivated by the fact that $\lambda_J$ depends non-linearly 
on the entire  thermal history of the IGM (see eqn.~\ref{eqn:jeans}), and both this
dependence and the thermal history are not well understood, as discussed in the introduction.
Allowing $\lambda_J$ to vary independently is the most straightforward
parametrization of our ignorance.  However, improvements in our theoretical understanding of the relationship
between $\lambda_J$ and the thermal history of the IGM ($T_0$,$\gamma$) could inform more intelligent parametrizations. 
Furthermore, inter-dependencies between thermal parameters can also be trivially included into our 
Bayesian
methodology for estimating the Jeans scale as conditional priors, e.g. $P(\lambda_J,T_0)$, in the parameter space.

\section{Power Spectra and Their Degeneracies}
\label{ps_cps}
Although many different statistics have been employed to isolate and
constrain the thermal information contained in Ly$\alpha$ forest
spectra, the flux probability density function (PDF; 1-point function) and
the flux power spectrum or auto-correlation function (2-point
function), are among the most
common\citep[e.g.][]{McDonald2000,Zald01,kbv+07,VielBolton09}.  But because
the Ly$\alpha$ transmission $F$ is significantly non-Gaussian,
significant information is also contained in higher-order statistics.
For example wavelet decompositions, which contains a hybrid of
real-space and Fourier-space information, have been advocated for
measuring spatial temperature fluctuations
\citep{Lidz09,Zaldarriaga02,Garzilli2012}. Several studies have
focused on the on the $b$-parameter distribution to obtain constraints
on thermal parameters
\citep{Ricotti00,Schaye00,McDonald2001,Rudie2012}, and recently
\citet{BeckerBolton2011} introduced a `curvature' statistic as an
alternative measure of spectral smoothness to the power spectrum.

As gas pressure acts to smooth the baryon density field in 3D, it is
natural explore power spectra as a means to constrain the Jeans
filtering scale. A major motivation for working in Fourier space, as
opposed to the real-space auto-correlation function, is that it is
much easier to deal with limited spectral resolution in Fourier
space. The vast majority of close quasar pairs are too faint to be
observed at echelle resolution ${\rm FWHM} \simeq 5\,{\rm km\,s^{-1}}$
where the Ly$\alpha$ forest is completely resolved. Instead, spectral
resolution has to be explicitly taken into account. 
But to a very good approximation the smoothing caused by limited spectral resolution simply low-pass
filters the flux, and thus the shape of the flux power spectrum is
unchanged for $k$-modes less than the spectral resolution cutoff
$k_{\rm res}$. Thus by working in $k$-space, one can simply ignore
modes $k \gtrsim k_{\rm res}$ and thus obviate the need to precisely
model the spectral resolution, which can be challenging for
slit-spectra. Finally, another advantage to $k$-space is that, 
because fluctuations in the IGM are only mildly non-linear, some of
the desirable features of Gaussian random fields, such as the
statistical independence of Fourier modes, are approximately retained,
simplifying error analysis. In what follows we consider the impact of
Jeans smoothing on longitudinal power spectrum, as well as the
simplest 2-point function that can be computed from quasar pairs, the
cross-power spectrum.

\subsection{The Longitudinal Power Spectrum}
\label{1dps}

\begin{figure*}
  \centering
  \centerline{\epsfig{file=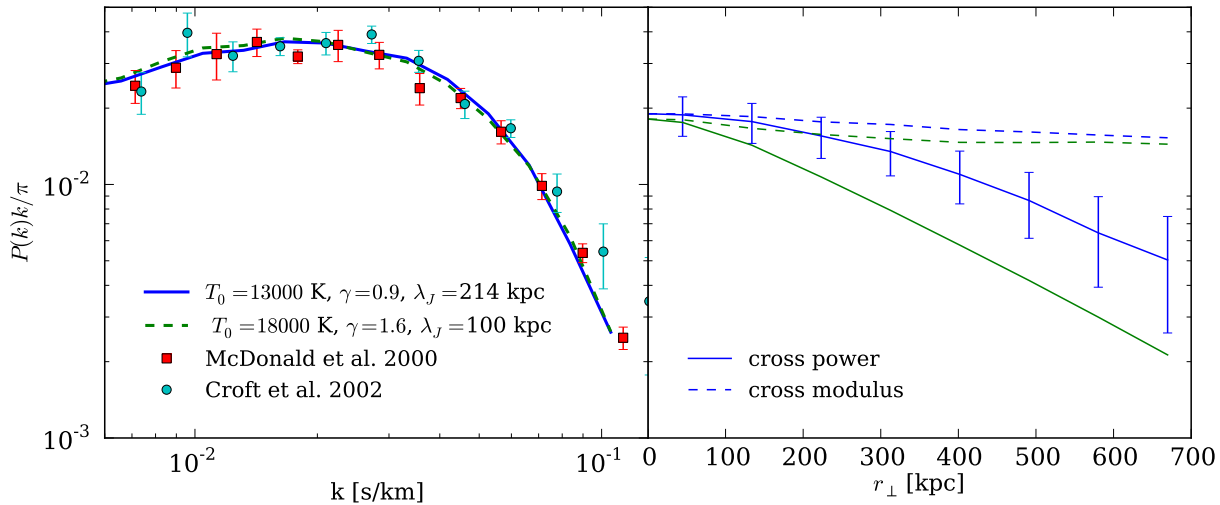,width=\textwidth}}
  \vskip -0.1in
\caption{ \label{fig:power_spectra}\emph{Left panel:} The 1D
  dimensionless power spectrum of the Ly$\alpha$ forest at $z=3$.  In our large
  grid of thermal models, we can identify two very different parameter
  combinations, represented by the solid (blue) and dashed (green)
  curves, which provide an equally good fit to the longitudinal power
  spectrum measurements from \citet{McDonald2000} (red squares) and
  \cite{Croft2002} (cyan circles), illustrating the strong
  degeneracies between these parameters
  ($T_0$,$\gamma$,$\lambda_J$). In light of these degeneracies, it is
  clear that it would be extremely challenging to constrain these
  parameters with the longitudinal power alone. \emph{Right panel:}
  The dimensionless cross power spectrum $\pi(k;r_{\perp})k/\pi$
  (solid line) at $k\approx 0.05$ s/km from our simulated skewers, as
  a function of $r_{\perp}$ for the same two thermal models shown at
  left, with error bars estimated from a sample of 20 pairs.  The
  degeneracy afflicting the 1D power is broken using the new
  information provided by close quasar pairs, because the different
  Jeans scales result in differing amounts of transverse spectral
  coherence, providing much better prospects for measuring
  $\lambda_J$.  We also show the cross modulus
  $\langle\rho_1(k)\rho_2(k)\rangle k/\pi$ (dashed lines) for the same
  two models, which show flat variation with $r_{\perp}$, and a very
  weak dependence on the Jeans scale. Most of the information about
  the 3D Jeans smoothing resides not in the moduli, but rather in the
  phase differences between homologous modes (see discussion in
  \S~\ref{sec:den}).}
\end{figure*}

It is well known that the shape of the longitudinal power spectrum,
and the high-$k$ thermal cutoff in particular, can be used constrain the
$T_0$ and $\gamma$ \citep{Zald01,VielBolton09}. This cutoff arises because
thermal broadening smooths $\tau$ in redshift-space
(e.g. eqn.~\ref{eqn:tau}). In contrast to this 1D
smoothing, the Jeans filtering smooths the IGM in 3D, and it is exactly this confluence between 1D and 3D
smoothing that we want to understand \citep[see
  also][]{Peeples09a,Peeples09b}.  We consider the quantity
$\delta F(v)=(F-\bar{F})/\bar{F}$, where $\bar{F}$ is the mean
transmitted flux, and compute the power spectrum according to
\begin{equation}
P(k)=\langle|\delta\tilde{F}(k)|^2\rangle, 
\label{los_pow_def}
\end{equation}
where $\delta\tilde{F}(k)$ denotes the Fourier transform of $\delta F$ for longitudinal
wavenumber $k$, and 
angular brackets denote an suitable ensemble average (i.e. over our full sample of spectra). 

In Figure~\ref{fig:power_spectra} we compare two thermal models in our thermal parameter grid
to measurements of the longitudinal power spectrum of the Ly$\alpha$
forest at $z\simeq 3$ \citep{McDonald2000,Croft2002}. The blue (solid)
curve has a large Jeans scale $\lambda_{\rm J} = 214\,{\rm kpc}$, a
cooler IGM $T_0=13,000\,$K, and a nearly isothermal
temperature-density relation $\gamma = 0.9$, which is mildly inverted
such that voids are hotter than overdensities. Such isothermal or even
inverted equations of state could arise at $z\sim 3$ from \ion{He}{2}
reionization heating \citep{McQuinn09,Tittley07}, and recent
analyses of the flux PDF \citep{Bolton08} as well joint analysis of
PDF and power-spectrum \citep{VielBolton09, Calura2012, Garzilli2012} have argued
for inverted or nearly isothermal values of $\gamma$. The green
(dashed) curves have a smaller Jeans scale $\lambda_{\rm J} =
100\,{\rm kpc}$, a hotter IGM $T_0=18,000\,$K, and a steep $\gamma =
1.6$ temperature-density relation consistent with the asymptotic value
if the IGM has not undergone significant recent heating events
\citep{HuiGnedin97,HH03}.  Thus with regards to the
longitudinal power spectrum, the Jeans scale is clearly degenerate
with the amplitude and slope $(T_0,\gamma)$ of the temperature-density
relation. One would clearly come to erroneous conclusions about the
equation of state parameters ($T_0$,$\gamma$) from longitudinal power
spectrum measurements, if the lack of knowledge of the Jeans scale is not
marginalized out \citep[see e.g.][for an example of this
  marginalization]{Zald01}. 

This degeneracy in the longitudinal power arises because the Jeans
filtering smooths the power in 3D on a scale which project to a
longitudinal velocity
\begin{equation}
  v_J=\frac{H(z=3)}{1+3} \lambda_J \approx 26 \left(\frac{\lambda_J}{340 \mbox{ kpc}}\right)\,\kms,
\end{equation}
resulting in a cutoff of the power at $k_J\approx0.04\,\skm$ (for the
typical values assumed in the introduction\footnote{We caution that this
estimate assumes a thermal history where $T\propto 1+z$, without
considering the effect of HeII reionization. In that case the
deduced value for the filtering scale $\lambda_J$ would probably be smaller.}). 
The thermal Doppler broadening of Ly$\alpha$ absorption lines smooths the power in 1D, on a scale governed
by the \emph{b-parameter} 
\begin{equation}
  b =\sqrt{\frac{2 k_B T}{\mu m_p}}\approx 15.7 \left(\frac{T}{1.5\times 10^4 \mbox{ K}}\right)^{1/2}\,\kms, 
\end{equation}
which results in an analogous cutoff at $k_{\rm th} = \sqrt{2}\slash b
\approx 0.09\,\skm$ for a temperature of 15000 K. Above $k_B$ is the Boltzmann constant, $m_p$ the
proton mass, and $\mu\approx 0.59$ is the mean molecular weight for a
primordial, fully ionized gas. The fact that the two cutoff scales are
comparable results in a strong degeneracy which is very challenging to
disentangle with longitudinal observations alone. Similar degeneracies
between the Jeans scale and ($T_0$,$\gamma$) exist if one considers
wavelets, the curvature, the  $b$-parameter distribution, and the flux PDF, 
which we
explore in an upcoming study (Rorai et al. 2013, in prep). In the next
section we show that this degeneracy between 3D and 1D smoothing can
be broken by exploiting additional information in the transverse
dimension provided by close quasar pairs.

\subsection{Cross Power Spectrum}
\label{sec:ps_cps}

The foregoing discussion illustrates that the 3D (Jeans) and 
1D (thermal broadening) smoothing are mixed in the longitudinal power
spectrum, and ideally one would measure the full 3D power spectrum to
break this degeneracy.  For an isotropic random field the 1D power
spectrum $P(k)$ and the 3D power $P_{3D}(k)$ are related according to
\begin{equation}
 P_{3D}=\frac{1}{2\pi} \frac{1}{k}\frac{dP(k)}{dk} \label{eqn:3dpow}.
\end{equation}
However, in the Ly$\alpha$ forest redshift-space distortions and
thermal broadening result in an anisotropies that render this expression invalid. 

With close quasar pairs, transverse correlations measured across the
beam contain information about the 3D power, and can thus
thus disentangle the 3D and 1D smoothing. Consider for example the
cross-power spectrum $\pi(k,r_{\perp})$ of two spectra
$\delta F_1(v)$ and $\delta F_2(v)$ separated by a transverse distance $r_{\perp}$
\begin{equation}
  \pi(k;r_\perp) =\Re[\delta \tilde{F}^*_1(k)\delta\tilde{F}_2(k)].  
\label{cps_pow_def}
\end{equation}
When $r_{\perp} \rightarrow 0$ then $\delta F_2 \rightarrow
\delta F_1$ and the cross-power tends to the longitudinal power $P(k)$.  The
cross-power can be thought of as effectively a power spectrum in the
longitudinal direction, and a correlation function in the transverse
direction \citep[see also][]{Viel2002}.  Alternatively stated, the cross
power provides a transverse distance dependent correction to the
longitudinal power $P(k)$, reducing it from its maximal value at
`zero lag' $r_{\perp}=0$.  This further implies that measuring the
cross power of closely separated and thus highly coherent spectra
amounts to, at some level, a somewhat redundant measurement of the
longitudinal power which could be simply deduced from isolated
spectra. In the next section, we will explain how to isolate the
genuine 3D information provided by close quasar pairs using a
statistic that is more optimal than the cross-power. Nevertheless,
Figure~\ref{fig:power_spectra} shows the cross-power spectrum for the 
two degenerate models discussed in the previous section, clearly 
illustrating that even the sub-optimal
cross-power spectrum can break the strong degeneracies between thermal
parameters that are present if one considers the longitudinal 
power alone.

\section{Phase Angles and the Jeans Scale}\label{phase_ang}

Although the cross-power has the ability to break the degeneracy
between 3D and 1D smoothing present in the longitudinal power,
we demonstrate here that the cross-power (or equivalently the cross-correlation
function) is however not optimal, and indeed the genuine 3D
information is encapsulated in the \emph{phase differences} between
homologous Fourier modes.

\subsection{Drawbacks of the Cross Power Spectrum}
\label{cps_vs_phase}
Let us write the 1D Fourier transform of the field $\delta F$ as
\begin{equation}
\delta \tilde{F}(k) = \rho (k) e^{i\theta(k)}
\end{equation}
where the complex Fourier coefficient is described by a modulus $\rho$ and phase angle 
$\theta$, both of which depend on $k$. Note that for any ensemble of spectra $P(k)=\langle \rho^2(k)\rangle$, hence 
the modulus $\rho(k)$ is a random draw from a distribution whose variance is given by 
the power spectrum.  From eqn.~(\ref{cps_pow_def}), the 
cross-power of the two spectra $\delta F_1(v)$ and  $\delta F_2(v)$ is then
\begin{equation}
\label{cpeq}
\pi_{12}(k) =\rho_1(k)\rho_2(k) \cos(\theta_{12}(k)), 
\end{equation}
where $\theta_{12}(k)=\theta_1(k)-\theta_2(k)$ is the phase difference
between the homologous $k-$modes.  The distribution of the moduli
$\rho_1$ and $\rho_2$ are also governed by $P(k)$, but at small impact
parameter they are not statistically independent because of spatial
correlations. Nevertheless, the moduli contain primarily information
already encapsulated in the longitudinal power, and are thus affected
by the same thermal parameter degeneracies that we described in the
previous section. For the purpose of constraining the Jeans scale, we
thus opt to ignore the moduli $\rho_1$ and $\rho_2$ altogether, in an
attempt to isolate the genuine 3D information, increasing sensitivity
to the Jeans scale, while minimizing the impact of thermal
broadening, removing degeneracies with the temperature-density relation
parameters ($T_0$,$\gamma$). 

The foregoing points are clearly illustrated by the dashed curves in 
the right panel of Figure~\ref{fig:power_spectra}, which compares the quantity $\langle
\rho_1(k)\rho_2(k)\rangle$ as a function of impact parameter $r_\perp$
for the same pair of thermal models discussed in \S~\ref{1dps}, which
are degenerate with respect to the longitudinal power. The similarity
of these two curves reflects the degeneracy of the
longitudinal power for these two models, and one observes a flat trend with $r_{\perp}$ and a 
very weak dependence on the Jeans scale $\lambda_J$, substantiating our argument that the moduli contain
primarily 1D information.

As the moduli contain minimal information about the 3D power, we are
thus motivated to explore how the phase difference $\theta_{12}(k)$
can constrain the Jeans scale. In terms of Fourier coefficients,
$\theta_{12}(k)$ can be written
\begin{equation}
 \theta_{12}(k)=\arccos\left(\frac{\Re[\delta \tilde{F}^*_1(k)
\delta \tilde{F}_2(k)]}{\sqrt{|\delta \tilde{F}_1(k)|^2|\delta \tilde{F}_2(k)|^2}}\right).
\label{eqn:phase}
\end{equation}
Note that because the phase difference is given by a ratio of Fourier
modes, it is completely insensitive to the normalization of $\delta
F$, and hence to quasar continuum fitting errors, provided that these
errors do not add power on scales comparable to $k$. In the remainder
of this section, we provide a statistical description of the
distribution of phase differences and we explore the properties and
dependencies of this distribution.
To simplify notation we will omit the subscript and henceforth denote the
phase difference as simply $\theta(k,r_{\perp})= \theta_1(k)-\theta_2(k)$,
where $r_{\perp}$ is the transverse distance between the two spectra 
$\delta F_1(v)$ and  $\delta F_2(v)$.

\subsection{An Analytical Form for the PDF of Phase Differences}
\label{sec:WC}
\begin{figure}
  \centering
  \centerline{\epsfig{file=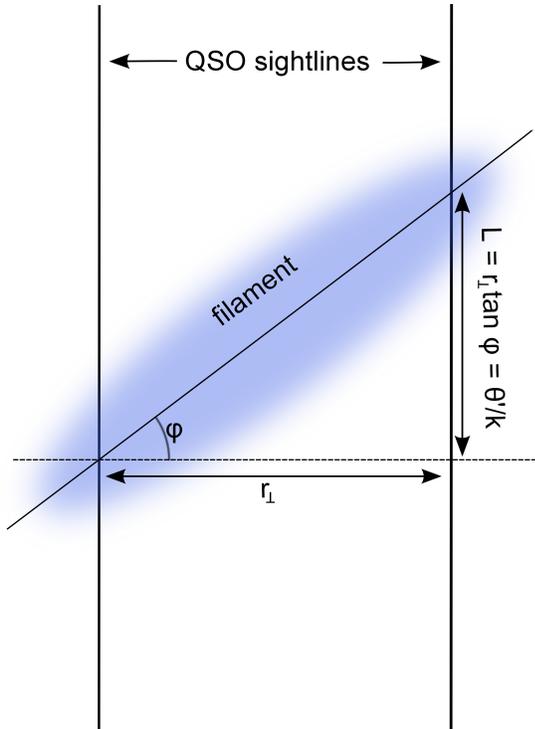,
      width=.4\textwidth}}
  \vskip -0.1in

  \caption{ \label{fig:ph_sc} Schematic representation of the
    heuristic argument used to determine the phase difference
    distribution: phase are determined by density filaments crossing
    the lines of sight of two quasars. If the orientation of the
    filaments $\varphi$ is isotropically distributed then $\theta^\prime$,
    dependent on the longitudinal distance $L=r_{\perp}\tan\varphi$,
    follows a Cauchy distribution.}
\end{figure}

The phase difference between homologous $k$-modes is a random variable
in the domain $[-\pi,\pi]$, which for a given thermal model, depends
on two quantities: the longitudinal mode in
question $k$ and the transverse separation $r_{\perp}$.  One might
advocate computing the quantity $\langle
\cos\theta(k,r_\perp)\rangle$ analogous to the cross-power (see
eqn.~\ref{cps_pow_def}), or the mean phase difference $\langle
\theta(k,r_\perp)\rangle$, to quantify the coherence of quasar pair
spectra.  However, as we will see, the
distribution of phase differences is not Gaussian, and hence is not
fully described by its mean and variance. This
approach would thus fail to exploit all the information encoded in its
shape.  Our goal is then to determine the functional form of the 
distribution of phase differences at any $(k,r_{\perp})$, and relate this to the thermal
parameters governing the IGM. This is a potentially daunting task, 
since it requires deriving a
unique function in the 2-dimensional space $\theta(k,r_{\perp})$
for any location in our 3-dimensional thermal parameter grid
$(T_0,\gamma,\lambda_J)$. Fortunately, we are able to reduce the complexity
considerably by deriving a simple analytical form for the
phase angle distribution.

We arrive at a this analytical form via a simple heuristic argument,
whose logic is more intuitive in real space. Along the same lines, we
focus initially on the IGM density distribution along 1D skewers, and
then later demonstrate that the same form also applies to the
Ly$\alpha$ flux transmission. Consider a filament of the cosmic web
pierced by two quasar sightlines separated by $r_\perp$, and oriented
at an angle $\varphi$ relative to the transverse direction.  A
schematic representation is shown in Figure~\ref{fig:ph_sc}.  This
structure will result in two peaks in the density field along the two
sightlines, separated by a longitudinal distance of $L=r_{\perp} \tan\varphi$. 
If we assume that the positions of these density maxima dictate the 
position of wave crests in Fourier space, the phase difference for 
a mode with wave  number $k$ can be written as
 $\theta^{\prime}=kL = k r_{\perp} \tan \varphi$. We can derive the
probability distribution of the phase difference by requiring that
$p(\theta^\prime)d\theta^\prime=p(\varphi)d\varphi$, and assuming
that, by symmetry, $\varphi$ is uniformly distributed. This implies
that $\theta^\prime$ follows the Cauchy-distribution
\begin{equation}
 p(\theta^\prime)=\frac{1}{\epsilon \pi }\frac{1}{1+ (\theta^\prime/\epsilon )^2}, 
\end{equation}
where $\epsilon$ parametrizes the distribution's concentration. 
As a final step, we need to
redefine the angles such that they reside in the proper
domain. Because $\tan \varphi$ spans the entire real line, so will
$\theta^\prime$; however, for any integer $n$, 
all phases $\theta^{\prime} + 2\pi n$
corresponding to distances $L + 2\pi n\slash k$ will map to
identical values of $\theta$, defined to be the phase difference in the domain 
$[-\pi,\pi]$. Redefining
the domain, requires that we re-map our probabilities according to 
\begin{equation}
  P_{[-\pi,\pi]}(\theta)=\sum_{n\in \mathbb{Z}}p(\theta + 2\pi n), 
\end{equation}
a procedure known as `wrapping' a distribution. Fortunately, the exact form of the 
wrapped-Cauchy distribution is known:
\begin{equation}
P_{\rm WC}(\theta)= \frac{1}{2\pi}\frac{1-\zeta^2}{1+ \zeta^2 - 2\zeta \cos(\theta - \mu)}, 
\label{WCD}
\end{equation}
where $\mu=\langle\theta\rangle$ is the mean value (in our case
$\mu=0$ by symmetry), and $\zeta$ is a concentration parameter between
0 and 1, which is the wrapped analog of $\epsilon$ above.  In the
limit where $\zeta \rightarrow 1$ the distribution tends to a Dirac
delta function $\delta_D(x)$, which is the behavior expected for
identical spectra. Conversely, $\zeta=0$ results in a uniform
distribution, the behavior expected for uncorrelated spectra.
A negative $\zeta$ gives distributions peaked at $\theta = \pi$ and is
unphysical in this context. 

\subsection{The Probability Distribution of Phase Differences of the IGM Density}
\label{sec:den}

\begin{figure*}
    \psfrag{r= 71 kpc}[c][][1.5]{$r_{\perp}= 71$ kpc}
    \psfrag{r=142 kpc}[c][][1.5]{$r_{\perp}=142$ kpc}
    \psfrag{r=333 kpc}[c][][1.5]{$r_{\perp}=333$ kpc}
    \psfrag{r=666 kpc}[c][][1.5]{$r_{\perp}=666$ kpc}
  \centering \centerline{\epsfig{file=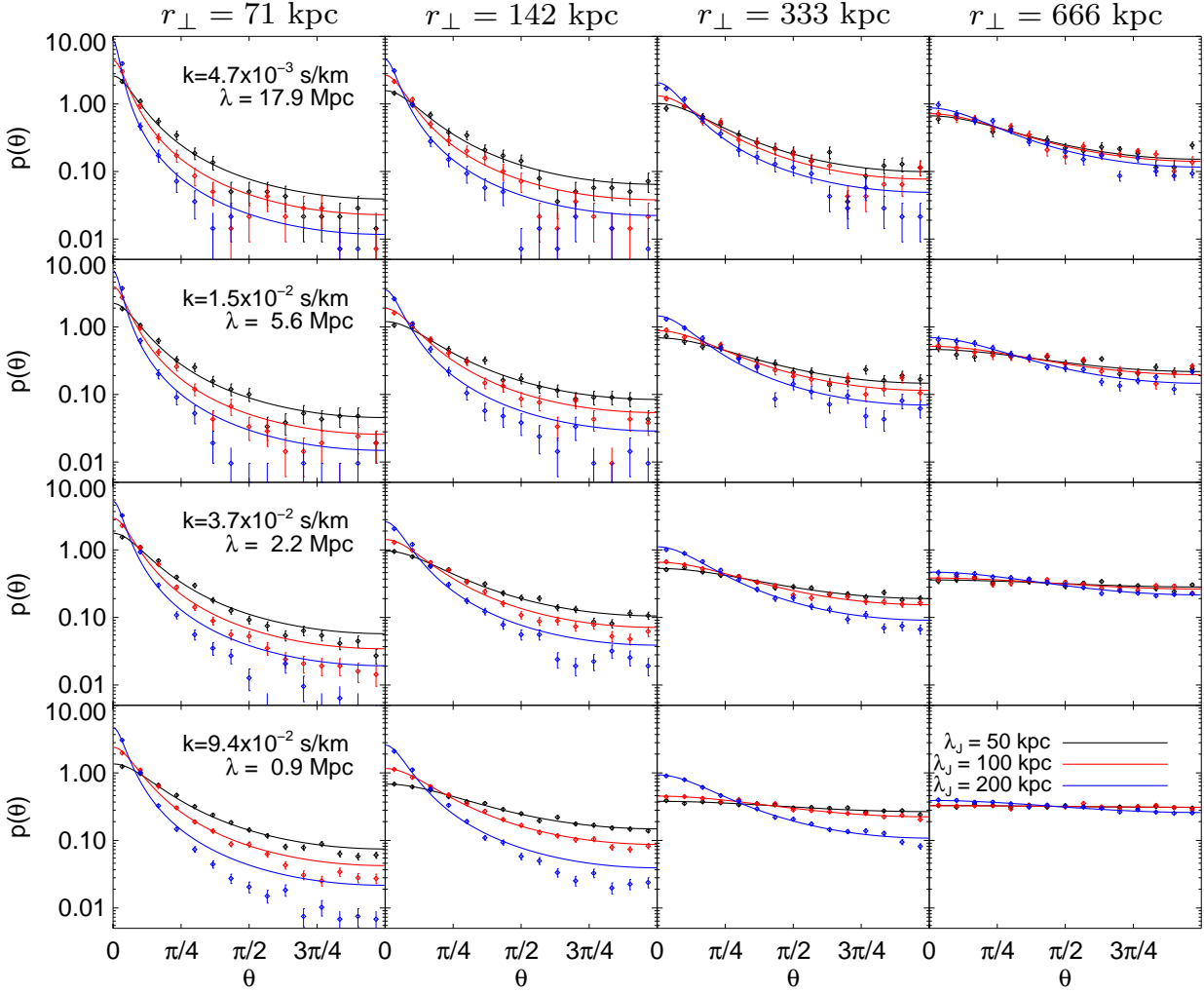,
      width=\textwidth}}
  \vskip -0.1in
  \caption{ \label{fig:denphase} Phase difference probability
    functions of the density fields at different
    separations $r_{\perp}$, wavenumbers $k$ and Jeans scale
    $\lambda_J$. Points with errorbars represent the binned phase
    distribution of the density field as obtained from the simulation,
    while the solid lines are the best-likelihood fit using a
    wrapped-Cauchy distribution. When the spectra are highly correlated
    the phases are small and the distribution is peaked around zero,
    whereas independent skewers result in flat probability functions.
    The error are estimated from the number of modes available 
    in the simulation, assuming a Poisson distribution.
    By symmetry $p(\theta)$ must be even in
    $\theta$, hence it is convenient to plot only the range $[0,\pi]$, summing
    positive and negative probabilities (clearly obtaining $p(|\theta|)$ ) to increase
    the sampling in each bin. We express the scale of each mode both giving
    the wavelength $\lambda$ in Mpc and the wave number $k$ (in s km$^{-1}$) in
    the transformed velocity space. The wrapped-Cauchy function traces with
    good approximation the phase distribution obtained from the simulation, showing
    less accuracy in the cases of strongly concentrated peaks, where low-probability
    bins are noisy. Each color is a different smoothing length: $\lambda_J=50,100$ and 200
    kpc (respectively black, red and blue). It is important to notice that the relative 
    distributions are different not only at scales comparable to $\lambda_J$, but also for 
    larger modes, because the 3D power of high-$k$ modes when projected on a 1D line 
    contributes to all the low-$k$ components (see the text for a detailed discussion). 
    Secondly, it is clear that the most relevant pairs are the closest ($r_{\perp} \lesssim \lambda_J$),
    because for wide separations the coherence is too low to get useful information. 
    These two consideration together explain why close quasar pairs are the most
    effective objects to measure the Jeans scale, even if they cannot be observed at
    high resolution.
    }
\end{figure*}

We now show that this wrapped-Cauchy form does a good job of
describing the real distribution of phase differences for our
simulated IGM density skewers.  Note that for our simple heuristic
example of randomly oriented filaments, the concentration parameter
$\zeta$ only depends on the product of $k r_{\perp}$; whereas, in the
real IGM, one expects the spectral coherence quantified by $\zeta$ to
depend on the Jeans scale $\lambda_J$. Because we do not know how to
directly compute the concentration parameter in terms of the Jeans
scale from first principles, we opt to calculate $\zeta$ from our
simulations. At any longitudinal wavenumber $k$, pair separation
$r_\perp$, and Jeans scale $\lambda_J$, our density skewers provide a
discrete sampling of the $\theta$ distribution. We use the maximum
likelihood procedure from \citet{Jammalamadaka} to calculate
the best-fit value of $\zeta$ from an ensemble of $\theta$ values, as
described further in Appendix \ref{rho_determin}. Figure \ref{fig:denphase} shows the
distribution of phases determined from our IGM density skewers
(symbols with error bars) compared to the best-fit wrapped-Cauchy
distributions (curves) for different longitudinal modes $k$,
transverse separations $r_\perp$, and values of the Jeans scale
$\lambda_J$. We see that the wrapped-Cauchy distribution typically
provides a good fit to the simulation data points to within the
precision indicated by the error bars. For very peaked distributions which
correspond to more spectral coherence (i.e. low-$k$ or large
$\lambda_J$), there is a tendency for our wrapped-Cauchy fits to
overestimate the probability of large phase differences relative to
the simulated data, although our measurements of the probability are
very noisy in this regime. We have visually inspected similar curves
for the entire dynamic range of the relevant $k$, $r_\perp$ and
$\lambda_J$, for which the shape of the wrapped-Cauchy distribution
varies from nearly uniform $(\zeta\simeq 0)$ to a very high degree of
coherence $(\zeta\simeq 1)$, and find similarly good agreement.

It is instructive to discuss the primary dependencies of the phase
difference distribution on wavenumber $k$, separation $r_\perp$, and
the Jeans scale $\lambda_J$ illustrated in Figure \ref{fig:denphase}.
At a fixed wavenumber $k$, a large separation relative to the Jeans
scale results in a flatter distribution of $\theta$, which approaches
uniformity for $r_\perp\gg \lambda_J$.  The distribution approaches
the fully coherent limit of a Dirac delta function for $r_\perp\ll
\lambda_J$, and the transitions from a strongly peaked distribution to
a uniform one occurs when $r_{\perp}$ is comparable to the Jeans scale
$\lambda_J$. We see that quasar pairs with transverse separations
$r_\perp$ $\lesssim 3 \lambda_J$, contain information about the Jeans
scale, whereas this sensitivity vanishes for larger impact parameters.
At fixed $r_\perp$, lower $k$-modes (i.e. larger scales) are more
highly correlated (smaller $\theta$ values) as expected, because
sightlines spaced closely relative to the wavelength of the mode
$kr_{\perp}\ll 1$, probe essentially the same large scale density
fluctuation.  Overall, the dependencies in Figure \ref{fig:denphase}
illustrate that there is information about the Jeans smoothing spread
out over a large range of longitudinal $k$-modes. Somewhat surprisingly,  even modes
corresponding to wavelengths $\gtrsim 100$ times larger than
$\lambda_J$ can potentially constrain the Jean smoothing.

This sensitivity of very large-scale longitudinal $k$-modes to a much
smaller scale cutoff $\lambda_J$  in the 3D power merits further
discussion. First, note that the range of wavenumbers typically probed by
longitudinal power spectra of the Ly$\alpha$ forest lie in the range
$0.005\,\skm < k < 0.1\,\skm$ (see Figure~\ref{fig:power_spectra}),
corresponding to modes with wavelengths $60\,\kms < v < 1250\,\kms$ or
$830\,{\rm kpc} < \lambda < 17\,{\rm Mpc}$. Here the low-$k$
cutoff is set by systematics related to determining the quasar
continuum \citep[see e.g.][]{Lee2012}, whereas the high-$k$ cutoff
is adopted to mitigate contamination of the small-scale power from
metal absorption lines \citep{McDonald2000}. In principle
high-resolution (echelle) spectra FWHM$=5\,\kms$ probe even higher
wavenumbers as large as $k\simeq 3$, however standard practice is to
only consider $k\lesssim 0.1$ in model-fitting \citep[see
  e.g.][]{Zald01}. Thus even the highest $k$-modes at our
disposable $k\simeq 0.1$ correspond to wavelengths $\simeq 830\,{\rm
  kpc}$ significantly larger than our expectation for the Jeans scale
$\sim 100\,$kpc. Furthermore, we saw in \S~\ref{1dps} that degenerate
combinations of the Jeans smoothing and the IGM temperature-density
relation can produce the same small-scale cutoff in the longitudinal
power. Thus both metal-line contamination and degeneracies with thermal broadening
imply that while it is extremely challenging to resolve the Jeans scale spectrally, the great advantage of close
quasar pairs is that they resolve the Jeans scale spatially, provided
they have transverse separations $r_{\perp}$ comparable to $\lambda_J$. We will
thus typically be working in the regime where $k\slash k_{\perp} \ll
1$, where we define $k_\perp \equiv x_0\slash aHr_\perp$, where
$aHr_\perp$ is the transverse separation converted to a velocity and
$x_0=2.4048$ is a constant the choice of which will become clear
below.

In this regime, it is straightforward to understand why the phase
differences between large-scale modes are nevertheless sensitive to
the Jeans scale. Consider the quantity $\langle
\cos{\theta(k,r_{\perp})}\rangle$, which is related to the cross-power
discussed in \S~\ref{cps_vs_phase}.  This `moment' of the phase angle PDF
can be written
\begin{equation}
\langle \cos\theta(k,r_{\perp})\rangle = \int_{-\pi}^{\pi} P(\theta(k,r_{\perp}))
\cos\theta(k,r_{\perp})d\theta\label{eqn:moment},    
\end{equation}
which tends toward zero for totally uncorrelated spectra
($P(\theta)=1\slash 2\pi$) and towards unity for perfectly correlated,
i.e. identical spectra ($P(\theta)=\delta_D(\theta)$)
spectra. Following the discussion in \S~\ref{cps_vs_phase}, we can
write \bea \pi(k,r_\perp) = \langle \rho_1(k)\rho_2(k) \cos\theta(k,r_{\perp})\rangle
&\approx& \\\langle \rho_1(k)\rho_2(k)\rangle\langle
\cos\theta(k,r_{\perp})\rangle\nonumber &\approx& P(k)\langle
\cos\theta(k,r_{\perp})\rangle, \eea where the first approximation is a
consequence of the approximate Gaussianity of the density
fluctuations, and the second from the fact that $\langle \rho_1
\rho_2\rangle\approx P(k)$ for $k\slash k_\perp \ll 1$, as
demonstrated by the dashed curves in the right panel of
Fig~\ref{fig:power_spectra}.  Thus we arrive at
\begin{equation}
\langle \cos\theta(k,r_{\perp})\rangle \approx \frac{\pi(k,r_\perp)}{P(k)} = 
\frac{\int_k^\infty dq q J_0(r_\perp\sqrt{q^2 - k^2})P_{\rm 3D}(q)}{\int_k^\infty dq q P_{\rm 3D}(q)},\label{eqn:costheta} 
\end{equation}
where $J_0$ is the cylindrical Bessel function of order zero. 
The numerator and denominator of the last equality in
eqn.~(\ref{eqn:costheta}) follow from the definitions of the
longitudinal and cross power for an isotropic 3D power spectrum
\citep[see e.g.][]{Lumsden1989,Peacockbook,Hui99,Viel2002}. The
denominator is the familiar expression for the 1D power expressed as a
projection of the 3D power. Note that 1D modes with wavenumber $k$
receive contributions from all 3D modes with wavevectors $\ge k$, which 
results simply from the geometry of observing a 3D field along a 1D
skewer. A long-wavelength (low-$k$) 1D longitudinal mode can be produced
by a short-wavelength (high-$k$) 3D mode directed nearly perpendicular to the 
line of sight \citep[see e.g.][]{Peacockbook}.  The numerator of eqn.~(\ref{eqn:costheta}) is
similarly a projection over all high-$k$ 3D modes, but because of the non-zero separation of the 
skewers the 3D power spectrum is now modulated by the cylindrical Bessel function $J_0(x)$. 
Because $J_0(x)$ is highly oscillatory, the primary contribution to this projection integral will come
from arguments in the range $0 < x < x_0$.  
Here $x_0=2.4048$ is the first zero of $J_0(x)$, which
motivates our earlier definition of $k_\perp \equiv x_0\slash
aHr_\perp$. For larger arguments $x$, the decay of $J_0(x)$ and its
rapid oscillations will result in cancellation and negligible
contributions. Thus for $k\slash k_\perp \ll 1$, we can finally write
\begin{equation}
\langle \cos\theta(k,r_{\perp})\rangle \approx 
\frac{\int_k^{k_{\perp}}dq q J_0(r_\perp\sqrt{q^2 - k^2})P_{\rm 3D}(q)}{\int_k^\infty dq q P_{\rm 3D}(q)}.\label{eqn:costheta2}
\end{equation}
This equation states that the average value of the phase difference
between homologous $k$ modes is determined by the ratio of the 3D
power integrated against a `notch filter' which transmits the range
$[k,k_\perp]$, relative to the total integrated 3D power over the full
range $[k,\infty]$. Hence phase angles between modes with wavelengths
$\gtrsim 100$ times larger than $\lambda_J$, are nevertheless
sensitive to the amount of 3D power down to scales as small as the
transverse separation $r_{\perp}$. This results simply from the
geometry of observing a 3D field along 1D skewers, because the power in 
longitudinal mode $k$ is actually dominated by the superposition of 3D power from much 
smaller scales $\gg k$. Provided that quasar pair separations resolve
the Jeans scale $r_{\perp}\sim \lambda_J$, even large scale modes
with $k \ll k_{\perp} \sim 1\slash \lambda_{J}$ are sensitive to 
the shape of the 3D power on small-scales,
which explains the sensitivity of low-$k$ modes to the Jeans scale in 
Figure \ref{fig:denphase}.

\begin{figure*}
    \psfrag{r= 71 kpc}[c][][1.5]{$r_{\perp}= 71$ kpc}
    \psfrag{r=142 kpc}[c][][1.5]{$r_{\perp}=142$ kpc}
    \psfrag{r=333 kpc}[c][][1.5]{$r_{\perp}=333$ kpc}
    \psfrag{r=666 kpc}[c][][1.5]{$r_{\perp}=666$ kpc}
  \centering \centerline{\epsfig{file=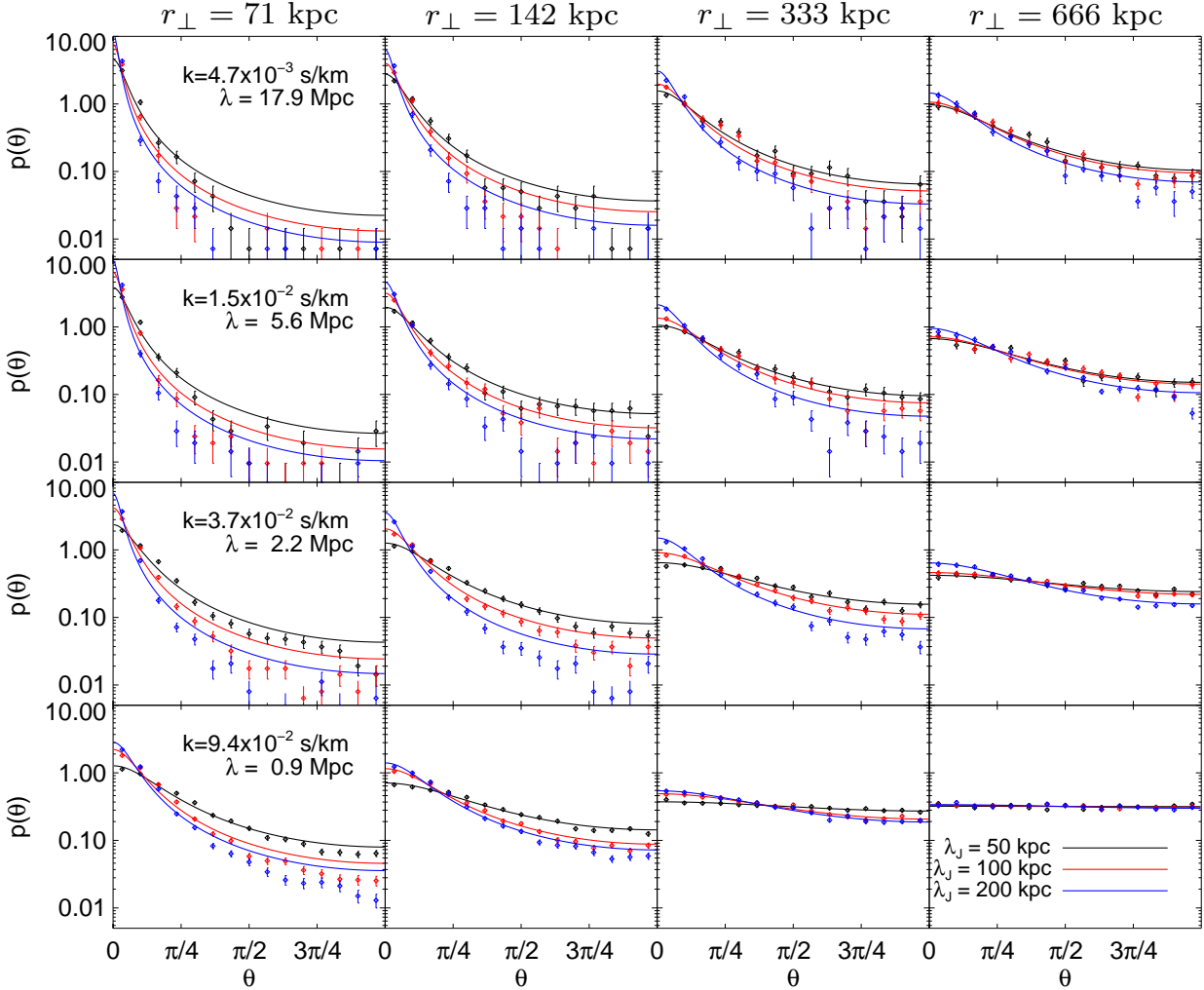,
      width=\textwidth}}
  \vskip -0.1in
  \caption{ \label{fig:fluxphase} Same plot of figure~\ref{fig:denphase} but
  for the \mlya\ transmitted flux field instead of density. We vary the Jeans 
  scale $\lambda_J$, keeping fixed the equation-of-state parameters, $T_0=10000$ K
  and $\gamma=1.6$. The properties of the distributions are analogous to the previous
  plot, they follow with good approximation a wrapped-Cauchy profile and they exhibit 
  the same trends with $r_{\perp},k$ and $\lambda_J$. Overall, the flux shows an higher degree
  of coherence and a slightly smaller sensitivity to $\lambda_J$.}
\end{figure*}

Finally, the form of eqn.~(\ref{eqn:costheta2}) combined with
eqn.~(\ref{eqn:moment}) explains the basic qualitative trends in
Figure~\ref{fig:denphase}. For large $r_{\perp}$ (small $k_{\perp}$)
the projection integral in the numerator decreases, $\langle
\cos\theta(k,r_{\perp})\rangle$ approaches zero, indicating that $P(\theta(k,r_{\perp}))$
approaches uniformity. Similarly, as $r_{\perp} \rightarrow
\lambda_J$, $\langle \cos\theta(k,r_{\perp})\rangle$ grows indicating that
$P(\theta(k,r_{\perp}))$ is peaked toward small phase angles, and in the limit
$r_{\perp} \ll \lambda_J$ $\langle \cos\theta(k,r_{\perp})\rangle \rightarrow
1$ and $P(\theta(k,r_{\perp}))$ approaches a Dirac delta function. At fixed
$r_{\perp}$, lower $k$ modes will result in more common pathlength in
the projection integrals in the numerator and denominator of
eqn.~(\ref{eqn:costheta2}), thus $\langle \cos\theta(k,r_{\perp})\rangle$ is
larger, $P(\theta(k,r_{\perp}))$ is more peaked, and the phase angles are more
highly correlated.

To summarize, following a simple heuristic argument, we derived a
analytical form for the phase angle distribution in \S~\ref{sec:WC}, 
which is
parametrized by a single number, the concentration $\zeta$. We
verified that this simple parametrization provides a good fit to the
distribution of phase differences in our simulated skewers, and
explored the dependence of this distribution on transverse separation
$r_{\perp}$, wavenumber $k$, and the Jeans scale $\lambda_J$. Phase
differences between large-scale modes with small wavenumbers $k \ll 1/\lambda_J$, are
sensitive to the Jeans scale, because geometry dictates that low-$k$
cross-power across correlated 1D skewers is actually dominated by
high-$k$ 3D modes up to a scale set by the pair separation $k_\perp
\sim 1\slash r_\perp$.

\subsection{The Probability Distribution of Phase Differences of the Flux}
\label{sec:flux}

\begin{figure*}
    \psfrag{r= 71 kpc}[c][][1.5]{$r_{\perp}= 71$ kpc}
    \psfrag{r=142 kpc}[c][][1.5]{$r_{\perp}=142$ kpc}
    \psfrag{r=333 kpc}[c][][1.5]{$r_{\perp}=333$ kpc}
    \psfrag{r=666 kpc}[c][][1.5]{$r_{\perp}=666$ kpc}
  \centering
  \centerline{\epsfig{file=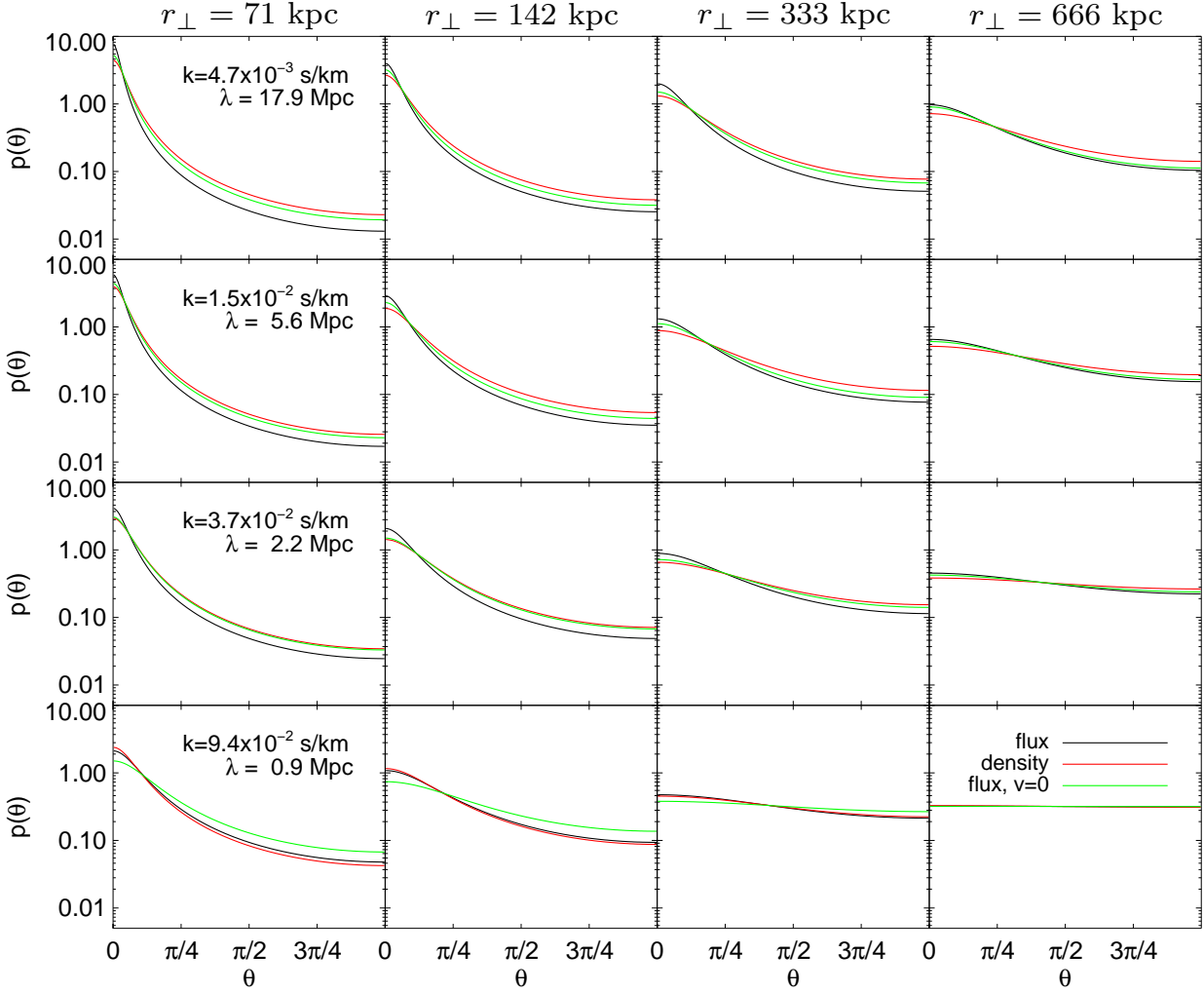,
      width=\textwidth}}
  \vskip -0.1in

  \caption{  \label{fig:fluxden}
Phase difference probability density functions for different separations
r⊥ and wavenumbers k. All models have the same Jeans scale λJ = 140 kpc. For clarity
we plot only the best-fit wrapped-Cauchy function without simulated points with
errorbars. The black and the red lines are the phase angle PDFs for the transmitted
flux of the Lyα forest and the IGM density field, respectively. 
The green line represents the case of the Lyα forest flux where peculiar velocities
are set to zero.  By comparing the green and the black lines we see that in peculiar
motions always increase the coherence between the two sightlines, which partly
explains the differences between the flux and density distributions, since the
latter is calculated in real space. The flux and density further differ because of
the non-linear FGPA transformation, which has a stronger effect on smaller scale
modes.}
\end{figure*}

\begin{figure*}
    \psfrag{r= 71 kpc}[c][][1.5]{$r_{\perp}= 71$ kpc}
    \psfrag{r=142 kpc}[c][][1.5]{$r_{\perp}=142$ kpc}
    \psfrag{r=333 kpc}[c][][1.5]{$r_{\perp}=333$ kpc}
    \psfrag{r=666 kpc}[c][][1.5]{$r_{\perp}=666$ kpc}
  \centering
  \centerline{\epsfig{file=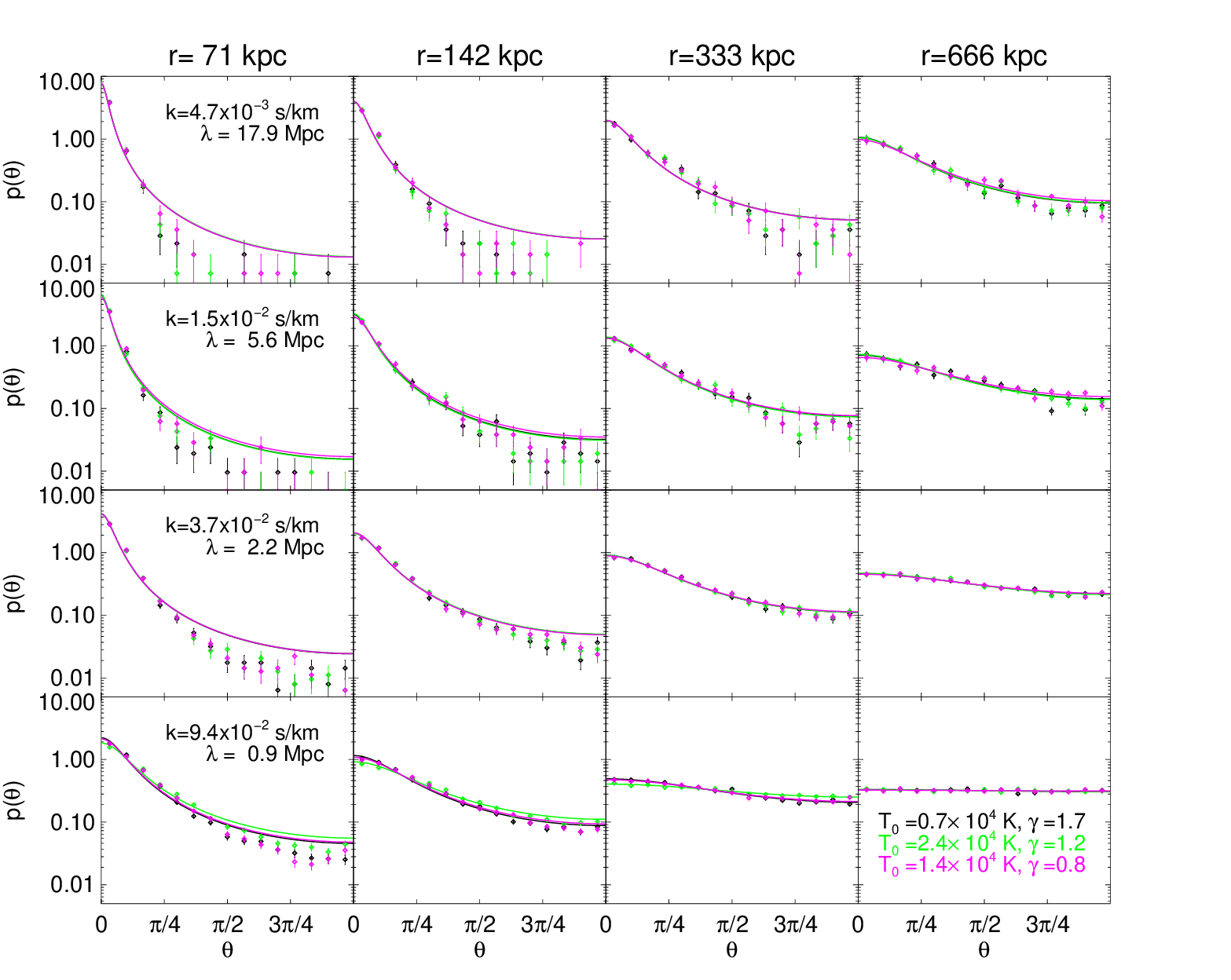,
      width=\textwidth}}
  \vskip -0.1in

  \caption{  \label{fig:ph_dist}Phase difference probability 
density functions for different separations $r_{\perp}$, wavenumbers $k$ and 
equation-of-state parameters $T_0 - \gamma$. Points with 
errorbars (estimated Poisson error) are the results of our simulations, 
while the coloured lines are the best-likelihood fit using 
a wrapped-Cauchy distribution. All models have the same Jeans 
scale $\lambda_J = 140$ kpc. This plot shows the most remarkable
property of phases: they do not exhibit any relevant 
sensitivity to the equation of state, so they 
robustly constrain the spatial coherence given by pressure 
support.}
\end{figure*}

Having established that the wrapped-Cauchy distribution provides a
good description of the phase difference of IGM density skewers, we
now apply it to the Ly$\alpha$ forest
flux. Figure~\ref{fig:fluxphase} shows the PDF of phase differences
for the exact same transverse separations $r_\perp$, wavenumbers $k$,
and Jeans smoothings $\lambda_J$ that were shown in
Figure~\ref{fig:denphase}. The other thermal parameters $T_0$ and
$\gamma$ have been set to $(T_0,\gamma)=(10,000\,{\rm K},1.6)$. Overall, the
behavior of the phase angle PDF for the flux is extremely similar to
that of the density, exhibiting the same basic trends. Namely, the
flux PDF also transitions from a strongly peaked distribution
($r_\perp\lesssim \lambda_J$) to a flat one ($r_\perp\gg \lambda_J$)
at around $r_\perp \simeq \lambda_J$.  Lower $k$-modes tend to be more
highly correlated, and low-$k$ modes corresponding to wavelengths
$\gtrsim 100 \lambda_J$ are nevertheless very sensitive to the Jeans
scale, in exact analogy with the density field. Note that because the 3D power spectrum 
of the flux field is now anisotropic, the assumptions leading to the derivation of 
eqn.~\ref{eqn:costheta2} in the previous section breaks down for the flux. 
Nevertheless, the explanation for the sensitivity of low-$k$ modes to the Jeans scale  
is likely the same, namely the
low-$k$ power across correlated skewers is actually dominated by projected
high-$k$ 3D power up to a scale $k_\perp \sim 1\slash r_\perp$, which is set 
by the pair separation.  

The primary difference between the phase angle PDF of flux versus the density
appears to be that the flux PDF is overall slightly less sensitive to
the Jeans scale. In general, we do not expect the two distributions to
be exactly the same for several reasons. 
First, the flux represents a highly nonlinear
transformation of the density: according to the FGPA formula $\delta F
\sim \exp{[-(1+\delta)^\beta]}$ where $\beta =
2-0.7(\gamma-1)$.  Second, the flux is observed in redshift space, 
and the peculiar velocities which determine the mapping from real to redshift
space, can further alter the flux relative to the density. 
Finally, the flux field is sensitive to other
thermal parameters $T_0$ and $\gamma$, both through the nonlinear FGPA
transformation, and because of thermal broadening. In what follows, we
investigate each of these effects in turn, and discuss how each alters
the phase angle PDF and its sensitivity to the Jeans scale.

In Figure~\ref{fig:fluxden} we show the flux PDF (black) alongside the 
density PDF (red) for various modes and separations, again with the
thermal model fixed to $(T_0,\gamma,\lambda_J)=(20,000 \mbox{
  K},1.0,140)$ kpc.  To isolate the impact of peculiar velocities, we
also compute the phase angle PDF of the real-space flux, i.e. without peculiar
velocities (green). Specifically, we disable peculiar velocities by
computing the flux from eqn.~(\ref{eqn:tau}) with $v_{p,\parallel}$
set to zero.  Overall, the PDFs of the real-space flux and density (also real-space) are quite
similar.  For low wavenumbers, the real-space flux skewers are always slightly more
coherent than the density ($P(\theta)$ more peaked) for all
separations. However, at the highest $k$, the situation is reversed
with the density being more coherent than the real-space flux.
A detailed explanation of the relationship between the phase angle PDF
of the real-space flux and the density fields requires a better understanding of
the effect of the non-linear FGPA transformation on the 2-point function
of the flux, which is beyond the scope of the present work.
Here we only argue that the 3D power spectrum of the real-space flux has in
general a different shape than that of the density, and using our
intuition from eqn.~(\ref{eqn:costheta2}), this will result in a
different shape for the distribution of phase angles. The net effect of peculiar velocities on 
the redshift-space flux PDF is to increase the amount of coherence 
between the two sightlines ($P(\theta)$ more peaked) 
relative to the real-space flux. This likely arises because the
peculiar velocity field is dominated by large-scale power, which 
makes the 3D power of the flux steeper as a function of $k$.  Again based
on our intuition from eqn.~(\ref{eqn:costheta}),  a steeper power
spectrum will tend to increase the coherence ($\langle \cos(\theta(k,r_{\perp}))\rangle$
closer to unity), because the projection integrals in
the numerator and denominator of eqn.~(\ref{eqn:costheta}) will both have 
larger relative contributions from the interval $[k,k_{\perp}]$. 
Note that the relative change in the flux PDF due to peculiar velocities is comparable to the
differences between the real-space flux and the density. At the highest $k$-values 
where the real-space flux is less coherent than the density (lowest
panel of Figure~\ref{fig:fluxden}), peculiar velocities conspire to make 
the redshift-space flux PDF very close to the density PDF.

Finally, we consider the impact of the other thermal parameters $T_0$
and $\gamma$ on the distribution of phases in
Figure~\ref{fig:ph_dist}. There we show the PDF of the phase
angles for the flux for a fixed Jeans scale $\lambda_J=140$\,kpc, and
three different thermal models.  Varying $T_0$ and $\gamma$ over the
full expected range of these parameters has very little impact on the
shape of the phase angle PDF, whereas we see in
Figure~\ref{fig:fluxphase} that varying the Jeans scale has a much
more dramatic effect. The physical explanations for the insensitivity
to $T_0$ and $\gamma$ are straightforward. The thermal parameters
$T_0$ and $\gamma$ can influence the phase angle PDF in two
ways. First, the FGPA depends weakly on temperature $T^{-0.7}$ through
the recombination coefficient. As a result the non-linear
transformation between density and flux depends weakly on $\gamma$
$\delta F \sim \exp{[-(1+\delta)^\beta]}$ where $\beta =
2-0.7(\gamma-1)$. We speculate that the tiny differences between the
thermal models in Figure~\ref{fig:ph_dist} are primarily driven by
this effect, because we saw already in Figure~\ref{fig:ph_dist} that
the non-linear transformation can give rise to large differences
between the density and flux PDFs. This small variation of the PDF
with $\gamma$ then suggests that it is actually the exponentiation
which dominates the differences between the flux and density PDFs in
Figure~\ref{fig:ph_dist}, with the weaker $\gamma$ dependent
transformation $(1+\delta)^{2-0.7(\gamma-1)}$ playing only a minor
role, which is perhaps not surprising. Note that there is also a
$T_0^{-0.7}$ dependence in the coefficient of the FGPA optical depth,
but as we require all models to have the same mean flux
$\langle \exp(-\tau)\rangle$, this dependence is compensated by the
freedom to vary the metagalactic photoionization rate $\Gamma$.
Second, both $T_0$ and $\gamma$ determine the temperature of gas at
densities probed by the Ly$\alpha$ forest, which changes the amount of
thermal broadening. The insensitivity to thermal broadening is also
rather easy to understand. Thermal broadening is effectively a
convolution of the flux field with a Gaussian smoothing kernel. In
$k$-space this is simply a multiplication of the Fourier transform of
the flux $\delta\tilde{F}(k)$ with the Fourier transform of the
kernel. Because all symmetric kernels will have a vanishing imaginary
part\footnote{The imaginary part of the Fourier transform of the
  symmetric function $W(|x|)$ is $\Im[W(k)]= \int W(|x|)\sin(kx)dx$
  which is always odd and will integrate to zero.}, the convolution
can only modify the moduli of the flux \emph{but the phases are
  invariant.} Thus the phase differences between neighboring flux
skewers are also invariant to smoothing, which explains the
insensitivity of the flux phase angle PDF to thermal broadening, and
hence the parameters $T_0$ and $\gamma$.

The results of this section constitute the cornerstones of our method for measuring
the Jeans scale. We found that the phase angle PDF of the flux has a shape very
similar to that of the density, and that both are well described by
the single parameter wrapped-Cauchy distribution. Information about
the 3D smoothing of the density field $\lambda_J$, is encoded in the
phase angle PDF of the flux, but it is essentially independent of the
other thermal parameters governing the IGM. This results because 1) 
the non-linear FGPA transformation is only weakly dependent on temperature
2) phase angles are invariant under
symmetric convolutions. The implication is that close quasar pair
spectra can be used to pinpoint the Jeans scale without suffering from
any significant degeneracies with $T_0$ and $\gamma$. Indeed, in
the next section we introduce a Bayesian formalism for estimating
the Jeans scale, and our MCMC analysis in \S~\ref{sec:measure} will
assess the accuracy with which the thermal parameters can be measured,
and explicitly demonstrate the near independence of constraints on
$\lambda_J$ from $T_0$ and $\gamma$.

\section{Estimating the Jeans Scale }\label{jeans_meas}

\begin{figure*}
    \psfrag{R1}[c][][1.5]{$r_{\perp}= 70$ kpc}
    \psfrag{R2}[c][][1.5]{$r_{\perp}=430$ kpc}
    \psfrag{log C}[c][][1.2]{$\log C_{\theta}$}
  \centering
   \centerline{\epsfig{file=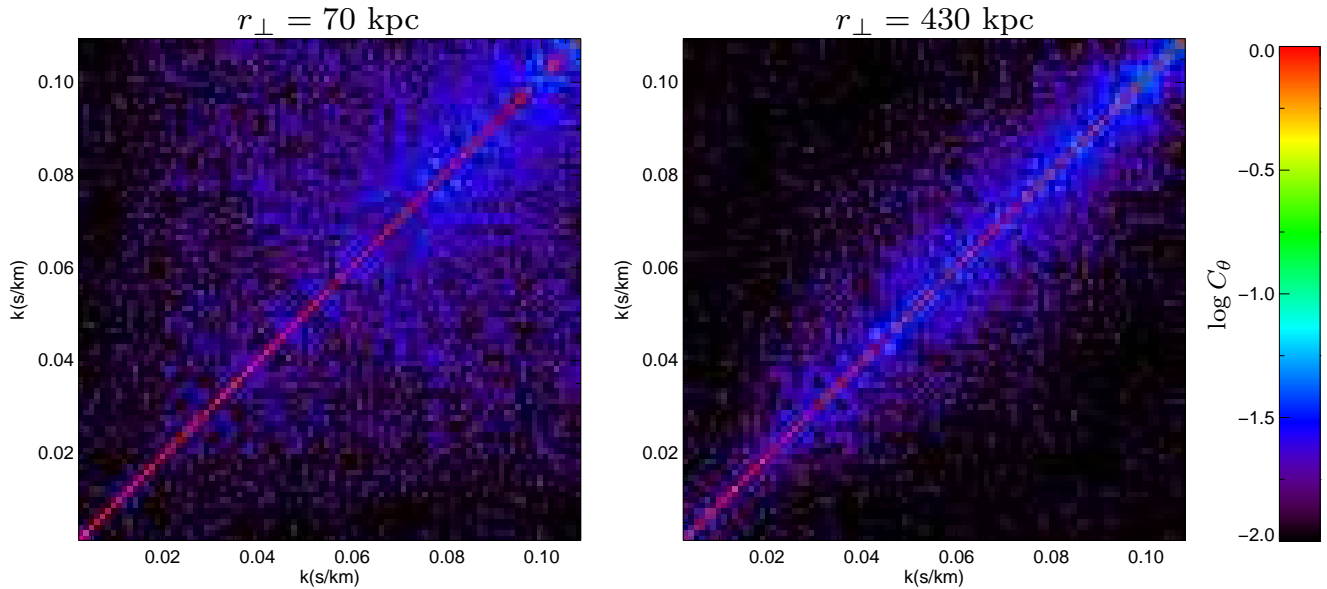,
       width=\textwidth}}
   \vskip -0.1in
  \caption{\label{fig:cov_mat}Logarithm of the phase $k-k$ correlation for separations $r_{\perp} = 70 $ kpc (left)
    and $r_{\perp}= 430$ kpc (right). This matrices are calculated for a model with $\lambda_J=143$ kpc, $T_0=20000$ K
    and $\gamma = 1$. Phases are more correlated when the impact parameter is smaller than the jeans
    scale and at high $k$ where nonlinear growth of perturbations couples different modes. Even in this 
    cases we rarely find correlations higher than $\approx 3\%$, for which reason we will work in 
    the diagonal approximation. This approximation may break out if the measured Jeans scale will be 
    significantly larger than expected.}
\end{figure*}

\subsection{The Covariance of the Phase Differences}
\label{sec:cov}

In the previous section, we showed that the PDF of 
phase differences between homologous longitudinal modes of the flux
field are well described by the wrapped-Cauchy distribution (see
eqn.~\ref{WCD}). However, the one-point function alone is
insufficient for characterizing the statistical properties of the
stochastic field $\theta(k,r_{\perp})$, because in principle values of
$\theta$ closely separate in either wavenumber $k$ or real-space could be
correlated. Understanding the size of these two-point correlations is of utmost
importance. Any given quasar pair spectrum provides us with a
realization of $\theta(k,r_{\perp})$, and we have seen that the
distribution of these values depends sensitively on the Jeans scale
$\lambda_J$. In order to devise an estimator for the thermal
parameters in terms of the phase differences, we have to understand
the degree to which the $\theta(k,r_{\perp})$ are independent.

It is easy to rule out the possibility of spatial correlations among
the $\theta$ values deduced from distinct quasar pairs. Because quasar
pairs are extremely rare on the sky, the individual quasar pairs in any
observed sample will typically be $\sim$ Gpc away from each other, and
hence different pairs will never probe correlated small-scale density
fluctuations. However, the situation is much less obvious when it
comes to correlations between $\theta$ values for different $k$-modes
of the same quasar pair. In particular, nonlinear structure formation
evolution will result in mode-mode coupling, which can induce correlations
between mode amplitudes and phases \citep[e.g.][]{Chiang2002,Watts2003,Coles2009}. We are
thus motivated to use our simulated skewers to directly quantity the
size of the correlations between phase differences of distinct
longitudinal $k$-modes.  

We calculate the correlation coefficient matrix of $\theta$ between
modes $k$ and $k^\prime$ defined as
\begin{equation}
C_{\theta}(k,k^\prime ; r_{\perp})=\frac{\left\langle \theta(k,r_{\perp})\theta(k^\prime,
r_{\perp}) \right\rangle}{\sqrt{\left\langle \theta^2(k,r_{\perp}) \right\rangle\left\langle\theta^2(k^\prime,r_{\perp})\right\rangle}}.  
\end{equation}
Our standard setup of $330$ pairs at each discrete separation $r_{\perp}$ results
in a very noisy estimate of $C_{\theta}(k,k^\prime;r_{\perp})$, so we proceed by
defining a new set of 80,000 skewers at two distinct discrete transverse
separations of $r_\perp =70$ kpc and $r_\perp = 430$ kpc for a 
single thermal model with $(T_0,\gamma,\lambda_J)=(20,000\,{\rm K},1,143\,{\rm
    kpc})$. 

Figure \ref{fig:cov_mat} displays the correlation
coefficient matrix for the two separations $r_{\perp}$ that we
simulated.  We find that the off-diagonal correlations between
$k$-modes are highest at high $k$ values and for smaller impact
parameters. This is the expected behavior, since higher
longitudinal $k$-modes will have a larger relative contributions from higher-$k$ 3D
modes, which will be more non-linear and have larger mode-mode correlations. 
Likewise, as per the discussion in \S~\ref{sec:den},
phase differences at smaller pair separations $r_\perp$ are sensitive to 
higher $k$ 3D power $\sim k_\perp$, and
should similarly exhibit larger correlations between modes. Note
however that over the range of longitudinal $k$ values which we will
use to constrain the Jeans scale $0.005 < k < 0.1$, the size of the
off-diagonal elements are always very small, of the order of $\sim 1-3\%$.

The small values of the off-diagonal elements indicates that the
mode-mode coupling resulting from non-linear evolution does not result
in significant correlations between the phase angles of longitudinal
modes. This could result from the fact that the intrinsic phase
correlations of the 3D modes is small, and it is also possible that
the projection of power inherent to observing along 1D skewers (see
\S~\ref{sec:den}) dilutes these intrinsic phase correlations, because a
given longitudinal mode is actually the average over a large range of
3D modes.  From a practical perspective, the negligible off-diagonal
elements in Figure~\ref{fig:cov_mat} are key, because
they allow us to consider each phase difference $\theta(k,r_{\perp})$
as an \emph{independent} random draw from the probability
distributions we explored in \S~\ref{sec:flux}, which as we show in
the next section, dramatically simplifies the estimator that we will
use to determine the Jeans scale.

\subsection{A Likelihood Estimator for the Jeans Scale}\label{sec:estimate}

The results from the previous sections suggest a simple method for
determining Jeans scale.  Namely, given any quasar pair, the phase angle
difference for a given $k$-mode represents a draw from the underlying
phase angle PDF determined by the thermal properties of the IGM (as
well as other parameters governing e.g. cosmology and the dark matter
which we assume to be fixed). In \S~\ref{sec:flux} we showed that the
phase angle PDF is well described by the wrapped-Cauchy distribution
and in \S~\ref{sec:cov} we argued that correlations between phase angle
differences $\theta(k,r_{\perp})$, in both $k$-space and real-space
can be neglected. Thus for a hypothetical dataset ${\theta(k,r_\perp)}$ measured from 
a sample of quasar pairs, we can write that the likelihood of the thermal model 
$M=\{T_0,\gamma,\lambda_J\}$ given the data is
\begin{equation}
 \mathscr{L}(\{\theta \}|M)=
 \prod_{i,j} P_{\rm WC}(\theta(k_i,r_j)|\zeta(k,r_\perp|M)).
\label{diaglik}
\end{equation}
This states that the likelihood of the data is the product of the phase angle PDF
evaluated at the measured phase differences for all $k$-modes and over
all quasar pair separations $r_\perp$.  Note that the simplicity of this
estimator is a direct consequence of the fact that there are
negligible $\theta$ correlations between different $k$-modes and pair
separations.  All dependence on $(T_0,\gamma,\lambda_J)$ is encoded in
the single parameter $\zeta$, which is the concentration of the
wrapped-Cauchy distribution (eqn.~\ref{WCD}). 

We can then apply Bayes' theorem to make inferences about any
thermal parameter, for example for $\lambda_J$
\begin{equation}
 P(\lambda_J|\{\theta \})=\frac{\mathscr{L}(\{\theta \}|\lambda_J) p(\lambda_J)}{P(\{\theta \})}
\label{fulllik}
\end{equation}
where $p(\lambda_J)$ is our prior on the Jeans scale and the
denominator acts as a renormalization factor which is implicitly
calculated by a Monte Carlo simulation over the parameter space. 
The same procedure can be used to evaluate the probability distribution of the
other parameters. Throughout this paper, we assume flat priors
on all thermal parameters, over the full domain of physically plausible parameter values. 

In \S~\ref{sec:measure} we will use MCMC techniques to numerically
explore the likelihood in eqn.~(\ref{fulllik}) and deduce the
posterior distributions of the thermal parameters. In order to do
this, we need to be able to evaluate the function
$\zeta(k,r_\perp|T_0,\gamma, \lambda_J)$ at any location in thermal
parameter space. This is a non-trivial computational issue, because we
do not have a closed form analytical expression for $\zeta$ which can be evaluated
quickly, and
thus have to resort to our cosmological simulations of the IGM to
numerically determine it for each model, as described in Appendix
\ref{rho_determin}. In practice, computational constraints limit the size of our 
thermal parameter grid to only 500 thermal models, and we 
thus evaluate $\zeta$ at only these 500 fixed locations. 
In the 
next section,  we describe a fast procedure referred to as an \emph{emulator}, 
which allows us to interpolate $\zeta$ from these 500 locations in our finite thermal 
parameter grid, onto any value in thermal parameter space $(T_0,\gamma,\lambda_J)$. 

\subsection{Emulating the IGM}
\label{sec:emulator}

Our goal is to define an algorithm to calculate
$\zeta(k,r_\perp|T_0,\gamma, \lambda_J)$ as a function of the thermal
parameters, interpolating from the values determined on a fixed
grid. As we will also compare Jeans scale constraints from the phase
angle PDF (eqn.~\ref{fulllik}), to those obtained from other
statistics, such as the longitudinal power $P(k)$ and cross-power
$\pi(k,r_\perp)$ (see \S~\ref{sec:measure}), we also need to be able
to smoothly interpolate these functions as well. To achieve this, we follow the
approach of the 'Cosmic Calibration Framework' (CCF) to provide an accurate
prediction scheme for cosmological observables
\citep{Heitmann06,Habib07}. The aim of the CCF is to build
\emph{emulators} which act as very fast -- essentially instantaneous
-- prediction tools for large scale structure observables such as the
nonlinear power spectrum \citep{Heitmann09,Heitmann2010,Lawrence09}, or
the concentration-mass relation \citep{Kwan12}.  Three essential steps
form the basis of emulation. First, one devises a sophisticated
space-filling sampling scheme that provides an optimal sampling
strategy for the cosmological parameter space being studied. Second, a
principle component analysis (PCA) is conducted on the measurements from the
simulations to compress the data onto a minimal set of basis functions
that can be easily interpolated. Finally, Gaussian process modeling is
used to interpolate these basis functions from the locations of the
space filling grid onto any value in parameter space. A detailed
description of our IGM emulator will be described in a companion paper
(A.Rorai et al. 2013, in preparation). Below we briefly summarize the key aspects.

Whereas CCF uses more sophisticated space filling Latin Hypercube
sampling schemes \citep[e.g.][]{Heitmann09}, we adopt a simpler
approach motivated by the shape of the IGM statistics we are trying to
emulate, which change rapidly at scales comparable to 
either the Jeans or thermal smoothing scale. We opt for an irregular
scattered grid which fills subspaces more effectively than a cubic lattice. 
We consider parameter values over the domain 
$\{(T_0,\gamma,\lambda_J):\,T_0 \in [5000,40000]\,{\rm K};\, \gamma \in [0.5,2];\, 
\lambda_J \in [43, 572]\,{\rm kpc}\}$. 
The lower limit of 43 kpc for the Jeans scale is chosen because this is about the smallest
value we can resolve with our simulation (see Appendix \ref{sec:appendixa}), while the
upper limit of 572 kpc is a conservative constraint deduced from the
longitudinal power spectrum: a filtering scale greater than this
value would be inconsistent with the high$-k$ cutoff, regardless of
the value of the temperature. The ranges considered for $T_0$ and $\gamma$ are 
consistent with those typically considered in the literature and our expectations
based on the physics governing the IGM. We sample the 3D thermal parameter
space at 500 locations, where we consider a discrete set of 50 points in each dimension. 
A linear spacing of these points is adopted for $\gamma$, whereas we find it 
more appropriate to distribute $T_0$ and $\lambda_J$ such that the scale of the cutoff 
of the power spectrum $k_{f}$ is regularly spaced. 
Since $k_f \propto \lambda_J^{-1}$ for Jeans smoothing and $k_f \propto T_0^{-1/2}$ 
for thermal broadening, we choose regular intervals of these parameters after transforming 
 $\lambda_J \rightarrow 1/\lambda_J$ and $T_0 \rightarrow 1/\sqrt{T_0}$. 
Each of the 50 values of the parameters is then repeated exactly 10 times in 
the 500-point grid, and we use 10 different random permutations of their indices  
to fill the space and to avoid repetition. For each thermal model in this grid, 
we generate 10,000 pairs of skewers at 30 linearly spaced discrete pair
separations between 0 and 714 kpc.  

We then use these skewers to compute the IGM statistics
$\zeta(k,r_\perp)$, $P(k)$, and $\pi(k,r_{\perp})$ for all $k$ and
$r_{\perp}$ for each thermal model. A PCA decomposition is then performed
in order to compress the information present in each statistic and represent its 
variation with the thermal parameters using a handful of basis functions $\phi$. 
A PCA is an orthogonal transformation that converts a family of correlated
variables into a set of linearly uncorrelated combinations of principal components. 
The components are ordered by the variance along each basis dimension, 
thus relatively few of them are sufficient to describe the entire
variation of a function in the space of interest, which is here the thermal parameter 
space.  To provide a concrete example, the longitudinal power spectrum $P(k)$ is fully 
described by the values of the power in each $k$ bin, but it is likely that some
of these $P(k)$ values do not change significantly given certain combinations of thermal 
parameters. The PCA determines basis functions of the $P(k)$ that best describe its variation
with thermal parameters, enabling us to represent this complex dependence
with an expansion onto just a few principal components
\begin{equation}
 P(k|T_0,\gamma,\lambda_J)=\sum_{i} \omega_{i}(T_0,\gamma,\lambda_J) \Phi_{i}(k),
\end{equation}
where $\{\Phi(k)\}$ are the basis of principal components, and
$\{\omega\}$ are the corresponding coefficients which depend on the
thermal parameters. The number of components for a given function is
set by the maximum tolerable interpolation errors of the emulator,
and these are in turn set by the size of the error bars on the
statistic that one is attempting to model.  We defer a detailed
discussion of the PCA analysis and the procedure used to determine the
number of components to an upcoming paper (Rorai et al. 2013, in
prep), but we note that the number of PCA components we used to fully represent the
functions $\zeta(k,r_\perp)$, $P(k)$, and $\pi(k,r_{\perp})$ were 25, 15, and 25, respectively
(phase distribution and cross power spectrum are 2D functions, so they need more components). 

Gaussian process interpolation is then used to interpolate these PCA
coefficients $\omega_{i}(T_0,\gamma,\lambda_J)$ from the irregular
distribution of points in our thermal grid to any location of interest
in the parameter space. The only input for the Gaussian interpolation
is the choice of \emph{smoothing length}, which quantifies the degree
of smoothness of each function along the direction of a given
parameter in the space.  We choose these smoothing lengths to be a
multiple of the spacing of our parameter grid. The choice of these
smoothing lengths is somewhat arbitrary, but we checked that the
posterior distributions of thermal parameters (eqn.~\ref{fulllik})
inferred do not change in response to a reasonable variations of these
smoothing lengths. A full description of the calibration and testing
of the emulator is presented in an upcoming paper (Rorai et al. 2013,
in prep).

To summarize, our method for measuring the Jeans scale of the IGM involves the following steps: 
\begin{itemize}

\item Calculate the phase differences ${\theta(k,r_\perp)}$ for each $k$-mode of an observed sample of quasar 
pairs with separations $r_\perp$. 

\item Generate Ly$\alpha$ forest quasar pair spectra for a grid of
  thermal models in the parameter space $(T_0,\gamma,\lambda_J)$, using
  our IGM simulation framework. For each model, numerically determine
  the concentration parameter $\zeta(k,r_\perp|T_0,\gamma, \lambda_J)$
  at each wavenumber $k$ and separation $r_{\perp}$, from 
  the distribution of phase differences $\theta(k,r_{\perp})$.

\item Emulate the function $\zeta(k,r_\perp|T_0,\gamma, \lambda_J)$,
  enabling fast interpolation of $\zeta$ from the fixed values in the
  thermal parameter grid to any location in thermal parameter space.

\item Calculate the posterior distribution in eqn.~(\ref{fulllik}) for $\lambda_J$, by exploring
  the likelihood function in eqn.~(\ref{diaglik}) with an MCMC algorithm. 

\end{itemize}


\section{How Well Can We Measure the Jeans Scale?}
\label{sec:measure}

Our goal in this section is to determine the precision with which
close quasar pair spectra can be used to measure the Jeans
scale. To this end, we construct a mock quasar pair dataset from our IGM simulations
and apply our new phase angle PDF likelihood formalism to it.  A key
question is how well constraints from our new phase angle technique
compare to those obtainable from alternative measures, such as the
cross-power spectrum, applied to the same pair sample, or from the
longitudinal power spectrum, measured from samples of individual
quasars. In what follows, we first present the likelihood used to
determine thermal parameter constraints for these two additional
statistics. Then we describe the specific assumptions made for the
mock data. Next we quantify the resulting
precision on the Jeans scale, explore degeneracies with other thermal
parameters, and compare to constraints from these two alternative statistics. We
explore the impact of finite signal-to-noise ratio and spectral resolution on our
measurement accuracy, and discuss possible sources of systematic error. Finally, we
explicitly demonstrate that our likelihood estimator provides unbiased
determinations of the Jeans scale.

\subsection{The Likelihood for $P(k)$ and $\pi(k,r_\perp)$}
\label{sec:p_lik}

For the longitudinal power $P(k)$, we assume that the distribution of
differences, between the measured band powers of a $k$-bin and the true
value, is a multi-variate Gaussian \citep[see e.g.][]{msb+06},
which leads to the standard likelihood for the power-spectrum
\bea
 \mathscr{L}(P_{d}|M) &=& (2\pi)^{-N\slash 2} \det{(\Sigma)}^{-1\slash 2} \\
 && \exp\left[-\frac{1}{2}(P_d-P_M)^T \Sigma^{-1}(P_d-P_M) \right]\nonumber, 
\label{P_los_lik}
\eea
where $P_d$ is a vector of $N$ observed 1D band powers, $P_M$ is a vector of power spectrum predictions 
for a given thermal model $M=(T_0,\gamma,\lambda_J)$, and 
\begin{equation}
\Sigma(k,k^{\prime}) = \langle [P(k)-P_M(k)][P(k^{\prime})-P_M(k^{\prime})]\rangle, 
\end{equation}
is the full covariance matrix of the power spectrum measurement. As we
describe in the next subsection, we will choose a subset of the
skewers from a fiducial thermal model to represent the `data' in this
expression, which are then compared directly to thermal models
$(T_0,\gamma,\lambda_J)$, where the same emulator technique described
in \S~\ref{sec:emulator} is used to interpolate
$P_M(k|T_0,\gamma,\lambda_J)$ to parameter locations in the thermal
space. To determine the covariance of this mock data
$\Sigma(k,k^{\prime})$, we use the full ensemble of $2\times 10,000$
1D skewers for the fiducial thermal model, directly evaluate the
covariance matrix, and then rescale it to the size of our mock dataset
by multiplying by the ratio of the diagonal terms
$\sigma^2_{\mbox{dataset}}/\sigma^2_{\mbox{full}}$.  This procedure of evaluating the
covariance implicitly assumes that the only source of noise in the
measurement is sample variance, or that the
instrument noise is negligible. For the high-resolution and high
signal-to-noise ratio spectra used to measure the longitudinal power
spectrum cutoff \citep{McDonald2000,Croft2002}, this is a reasonable
assumption. For reference, the relative magnitude of off-diagonal
terms of the covariance,
$\Sigma(k,k')/\sqrt{\Sigma(k,k)\Sigma(k',k')}$, are at most $20-30\%$
with the largest values attained at the highest $k$.

For the cross-power spectrum $\pi(k,r_\perp)$, we follow the same procedure.  Namely, a mock dataset
is constructed for the fiducial thermal model by taking a subset of the full ensemble of quasar pair
spectra. We again assume that the band power errors are distributed according to a multi-variate
Gaussian, but because we must now account for the variation with separation $r_{\perp}$, the corresponding 
likelihood is
\begin{equation}
 \mathscr{L}(\pi|M) = \prod_i \mathscr{L}(\pi_{d}(k,r_{\perp,i})|M)\label{eqn:crosslik}, 
\end{equation}
where $\mathscr{L}(\pi_{d}(k,r_{\perp,i})|M)$ has the same form as the longitudinal
power in eqn.~(\ref{P_los_lik}). In exact analogy with the longitudinal power,  we
compute the full covariance matrix $\Sigma(k,k^{\prime}|r_{\perp})$ of the cross-power using our full 
ensemble of simulated pair spectra for our fiducial model, but now at each value of $r_{\perp}$.

\subsection{Mock Datasets}
\label{sec:mock}
To determine the accuracy with which we can measure the Jeans scale
and study the degeneracies with other thermal parameters, we construct
a dataset with a realistic size and impact parameter distribution, and
use an MCMC simulation to explore the phase angle likelihood in
eqn.~(\ref{diaglik}). We compare these constraints to those obtained
from the cross-power spectrum for the same mock pair dataset, by
similarly using an MCMC to explore the cross-power likelihood in
eqn.~(\ref{eqn:crosslik}). We also compare to parameter constraints
obtainable from the longitudinal power alone, by exploring the
likelihood in eqn.~(\ref{P_los_lik}), for which we must also construct
a mock dataset for longitudinal power measurements.

For the mock quasar pair sample, we assume 20 quasar pair spectra at
$z=3$, with fully overlapping absorption pathlength between Ly$\alpha$
and Ly$\beta$. Any real quasar pair sample will be composed of both
binary quasars with full overlap and projected quasar pairs with
partial overlap, so in reality 20 represents the total effective pair
sample, whereas the actual number of quasar pairs required could be
larger. The distribution of transverse separations for these pairs is
taken to be uniform in the range $24 < r_{\perp} < 714$ kpc. Specifically, we require 200
pairs of skewers in order to build up the necessary path length for 20
full Ly$\alpha$ forests, and these are randomly selected from our
10,000 IGM pair skewers which have 30 discrete separations.  
We draw these pairs from a simulation with a
fiducial thermal model $(T_0,\gamma,\lambda_J)=(12,000\,{\rm K},1.0,
110,\,{\rm kpc})$, which lies in the middle of our
thermal parameter grid. Note that follow-up observations of
quasar pair candidates has resulted in samples of $> 400$ quasar
pairs in the range $1.6 < z \lesssim 4.3$ with $r_{\perp} < 700\,{\rm
  kpc}$, and for those with $> 50\%$ overlap, the total effective
number of fully overlapping pairs is $\simeq 300$
\citep{Hennawi04,BINARY,Myers08,HIZBIN}. Many of these sightlines
already have the high quality Ly$\alpha$ forest spectra required for a
Jeans scale measurement \citep[e.g.][]{QPQ1,QPQ2,QPQ3,QPQ4,QPQ5},
hence the mock dataset we have assumed already exists, and can be easily enlarged
given the number of close quasar pairs known. 

Longitudinal power spectrum measurements which probe the small-scale
cutoff of the power have been performed on high-resolution ($R\simeq
30,000-50,000$; FWHM=$6-10\,\kms$) spectra of the brightest quasars.
Typically, the range of wavenumbers used for model fitting is
$0.005\,\skm < k < 0.1\,\skm$ (see Figure~\ref{fig:power_spectra}),
where the low-$k$ cutoff is chosen to avoid systematics related to
quasar continuum fitting \citep{Lee2012}, and the high-$k$ cutoff is
adopted to mitigate contamination from metal absorption lines
\citep{McDonald2000,Croft2002,Kim04}. Because quasar pairs are very
rare, one must push to faint magnitudes to find them in sufficient
numbers. In contrast with the much brighter quasars used to measure
the small-scale longitudinal power
\citep{McDonald2000,Croft2002,Kim04}, quasar pairs are typically too
faint to be observable at echelle resolution (FWHM=$6-10\,\kms$) on 8m
class telescopes. However, quasar pairs can be observed with higher
efficiency echellette spectrometers, which deliver $R\simeq 10,000$ or
FWHM$=30\,\kms$. The cutoff in the power spectrum induced by this
lower resolution is $k_{\rm res}=1/\sigma_{\rm res} = 2.358/{\rm FWHM}
= 0.08\,\skm$, which is very close to the upper limit $k < 0.1\,\skm$
set by metal-line contamination. For these reasons, we will consider
only modes in the range $0.005\,\skm < k < 0.1\,\skm$ in the likelihood in
eqn.~(\ref{diaglik}). We initially consider perfect data, ignoring the
effect of finite signal-to-noise rate and resolution.  Then in
\S~\ref{sec:noise}, we will explore how noise and limited spectral resolution
influence our conclusions.

For the mock sample used to study the longitudinal power, 
we assume perfect data, which is reasonable considering 
that such analyses are typically performed on spectra with
signal-to-noise ratio ${\rm S\slash N}\sim 30$ and resolution FWHM$=6\,\kms$
\citep{McDonald2000,Croft2002,Kim04} such that the Ly$\alpha$ forest, and in particular
modes with $k< 0.1$, are fully resolved. For the size of this sample,
we again assume 20 individual spectra at $z=3$ with full coverage of
the Ly$\alpha$ forest, which is about twice the size employed in recently
published analyses \citep{McDonald2000,Croft2002,Kim04}. However, the number of
existing archival high-resolution quasar spectra at $z=3$ easily
exceeds this number, so samples of this size are also well within
reach. Also, adopting a sample for the longitudinal power with the same Ly$\alpha$
forest path length as the quasar pair sample, facilitates a
straightforward comparison of the two sets of parameter constraints.

\subsection{The Precision of the $\lambda_J$ Measurement}
\label{sec:accuracy}

\begin{figure*}
  \centering
  \centerline{\epsfig{file=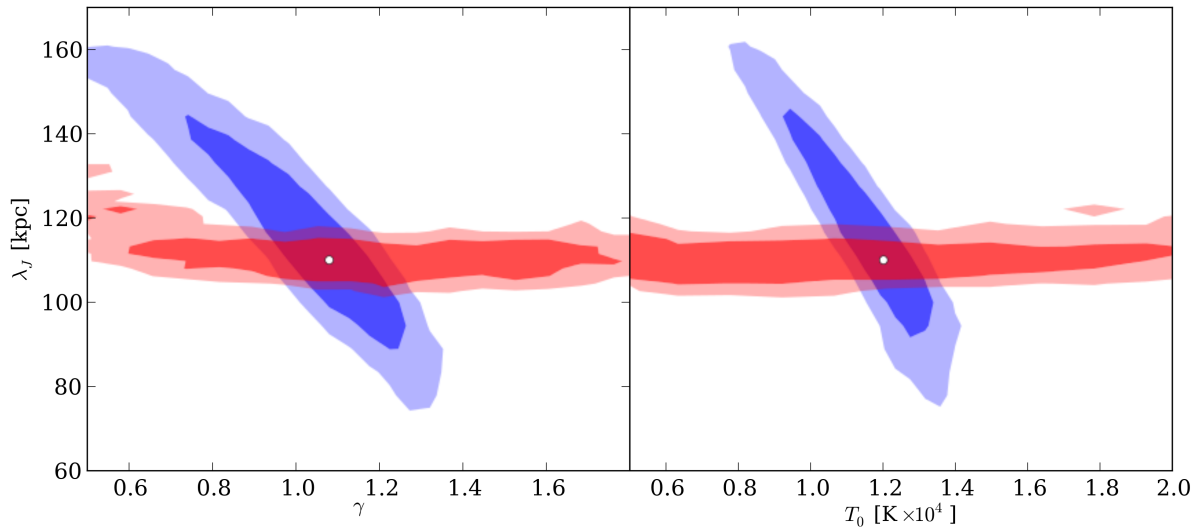,
      width=\textwidth}}
  \vskip -0.1in
  \caption{\label{fig:cont_phase}Constraints on the $\gamma-\lambda_J$ and $T_0-\lambda_J$ planes. The contours show
    the estimated $65\%$ and $96\%$ confidence levels
    obtained with the longitudinal power (blue) and the phase difference (red).
    The white dot marks the fiducial model in the parameter space.
    The degeneracy affecting the 1D power already shown in figure~\ref{fig:power_spectra}
    can now be seen clearly in the parameter space through the inclination of 
    the black contours. Conversely, the fact that constraints given by the phase difference
    statistic are horizontal guarantees that this degeneracy is broken and that 
    the measurement of the Jeans scale is not biased by the uncertainties on the 
    equation of state.}
\end{figure*}

Given our mock dataset and the expression for the phase angle likelihood in eqn.~(\ref{diaglik}), 
and armed with our IGM emulator, which enables us to quickly evaluate this likelihood 
everywhere inside our thermal parameter space, we are now ready to explore this likelihood
with an MCMC simulation to determine the precision with which we can measure the Jeans scale
and explore degeneracies with other thermal parameters. 

We employ the publicly available MCMC package described in
\cite{MCpackage}, which is particularly well adapted to explore
parameter degeneracy directions. The result of our MCMC simulation is
the full posterior distribution in the 3-dimensional
$T_0-\gamma-\lambda_J$ space for each likelihood that we consider. It
is important to point out that, in general, these posterior
distributions will not be exactly centered on the true fiducial
thermal model $(T_0,\gamma,\lambda_J)={12,000\,{\rm K}, 1, 110,\,{\rm
    kpc}}$.  Indeed, the expectation is that the mean or mode of the
posterior distribution for a given parameter will scatter around the
true fiducial value at a level comparable to the width of this
distribution. Nevertheless, the posterior distribution should provide
an accurate assessment of the precision with which parameters can be
measured and the degeneracy directions in the parameter space. In
\S~\ref{sec:bias} we will demonstrate that our phase angle PDF
likelihood procedure is indeed an unbiased estimator of the Jeans
scale, by applying this method to a large ensemble of
mock datasets, and showing that on average, we recover the input
fiducial Jeans scale.

The red shaded regions in Figure~\ref{fig:cont_phase} show the constraints
in thermal parameter space resulting from our MCMC exploration of the
phase angle likelihood (eqn.~\ref{diaglik}). The results are shown
projected onto the $T_0-\lambda_J$ and $\gamma-\lambda_J$ planes,
where the third parameter ($\gamma$ and $T_0$, respectively) has been
marginalized over. The dark and light shaded regions show $65\%$ and
$96\%$ confidence levels, respectively. The phase difference technique
(red) yields essentially horizontal contours, which pinpoint the value
of the Jeans scale, with minimal degeneracy with other thermal
parameters. This is a direct consequence of the near independence of
the phase angle PDF of $T_0$ and
$\gamma$ shown in Figure~\ref{fig:fluxden}, and discussed in
\S~\ref{sec:flux}.  The physical explanation for this independence is
that 1) the non-linear FGPA transformation in only 
weakly dependent on temperature 2) phase angles are invariant to the
thermal broadening convolution. This truly remarkable result is the
key finding of this work: phase angles of the Ly$\alpha$ forest flux
provide direct constraints on the 3D smoothing of the IGM density
independent of the other thermal parameters governing the IGM.

\begin{figure*}
  \centering
  \centerline{\epsfig{file=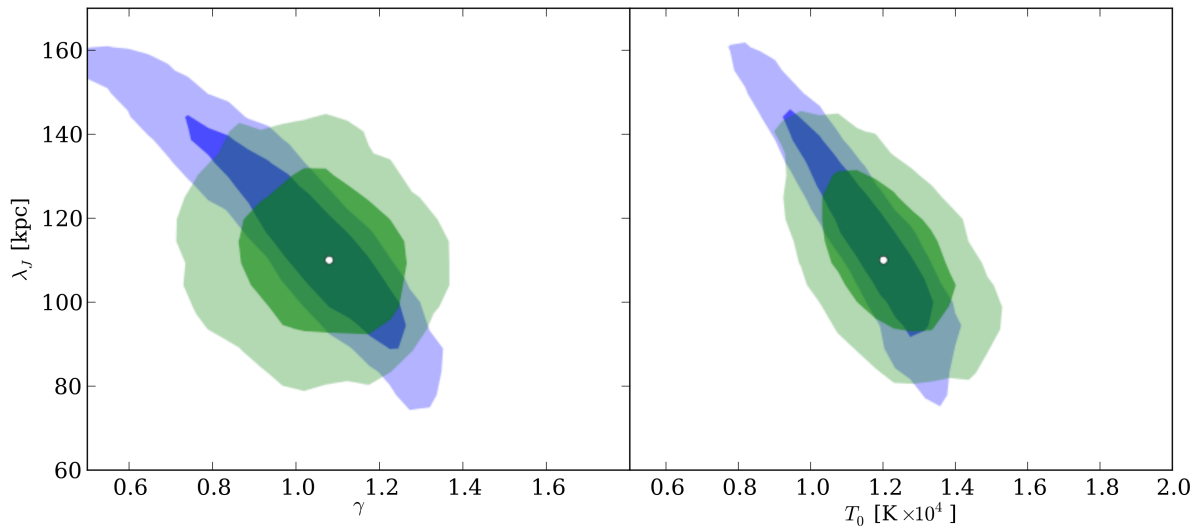,
      width=\textwidth}}
  \vskip -0.1in
  \caption{\label{fig:cont_cps}Constraints on the $\gamma-\lambda_J$ 
    and $T_0-\lambda_J$ planes. The contours show the
    estimated $65\%$ and $96\%$ confidence levels
    obtained with the longitudinal power (blue) and the cross power
    (green). The white dot marks the fiducial model in the parameter space. 
    Comparing this plot with figure~\ref{fig:cont_phase} makes clear why the 
    cross power spectrum is not the optimal statistic for measuring $\lambda_J$
    since the phase information is diluted and the degeneracy is not efficiently
    broken.
}
\end{figure*}

The blue shaded regions in Figure~\ref{fig:cont_phase} show the
corresponding parameter constraints for our MCMC of the longitudinal
power spectrum likelihood (eqn.~\ref{P_los_lik}). Considering the
longitudinal power spectrum alone, we find that significant
degeneracies exist between $\lambda_J$, $T_0$ and $\gamma$, which
confirms our qualitative discussion of these degeneracies in \S~\ref{1dps}
and illustrated in Figure~\ref{fig:power_spectra}. These
degeneracy directions are easy to understand. The longitudinal power
is mostly sensitive to thermal parameters via the location of the
sharp small-scale cutoff in the power spectrum. This thermal cutoff is set by a
combination of both 3D Jeans pressure smoothing and 1D thermal
broadening. The thermal broadening component is set by the temperature
of the IGM at the characteristic overdensity probed by the forest, which is
$\delta \approx 2$ at $z=3$ \citep{BeckerBolton2011}. One naturally
expects a degeneracy between $T_0$ and $\gamma$, because it is
actually the temperature at $T(\delta \approx 2)$ that sets the
thermal broadening. A degeneracy between $\lambda_J$ and $T(\delta
\approx 2)$ is also expected because both smoothings contribute to the
small-scale cutoff. Thus, a lower Jeans scale can be compensated by
more thermal broadening, which can result from either a steeper
temperature density relation (larger $\gamma$) or a hotter temperature
at mean density $T_0$, since both produce a hotter $T(\delta \approx
2)$.

Previous work that has aimed to measure thermal parameters such as
$T_0$ and $\gamma$, from the longitudinal power spectrum
\citep{Zald01,VielBolton09}, the curvature statistic
\citep{BeckerBolton2011}, wavelets \citep{Theuns02b,Lidz09,Garzilli2012}, 
and the  $b$-parameter distribution 
\citep{Haehnelt98,Theuns00,Ricotti00,BryanMach00,Schaye00,McDonald2001,Theuns02a,Rudie2012},
have for the
most part ignored the degeneracies between these thermal parameters
and the Jeans scale (but see Zaldarriaga et al. 2001 who marginalized
over the Jeans scale, and Becker et al. 2011 who also considered its
impact). Neglecting the possible variation of the Jeans scale is
equivalent to severely restricting the family of possible IGM thermal
histories.  Because the phase angle method accurately pinpoints the
Jeans scale independent of the other parameters, it breaks the
degeneracies inherent to the longitudinal power spectrum and will enable
accurate and unbiased measurements of both $T_0$ and $\gamma$, as
evidence by the intersection of the red and black contours in
Figure~\ref{fig:cont_phase}.  Similar degeneracies between the Jeans
scale and ($T_0$,$\gamma$) exist when one considers other statistics such 
as the flux PDF \citep{McDonald2000,kbv+07,Bolton08,Calura2012,Garzilli2012}, 
which we will explore in an upcoming study (Rorai et al. 2013, in prep). In light 
of these significant degeneracies with the Jeans scale, it may be necessary to reassess the
reliability and statistical significance of previous measurements of $T_0$ and $\gamma$.

Figure~\ref{fig:cont_cps} shows the resulting thermal parameter
constraints for our MCMC analysis of the cross-power spectrum
likelihood (eqn.~\ref{eqn:crosslik}) in green, determined from exactly
the same mock quasar pair sample that we analyzed for the phase
angles. The confidence regions for the longitudinal power are shown
for comparison in blue.  The cross-power spectrum is a straightforward
statistic to measure and fit models to, and the green confidence
regions clearly illustrate that it does exhibit some sensitivity to
the Jeans scale, as discussed in \S~\ref{sec:ps_cps} and shown in the
right panel of Figure~\ref{fig:power_spectra}. However, a comparison
of the cross-power confidence regions in Figure~\ref{fig:cont_cps}
(green) with the phase angle PDF confidence regions in
Figure~\ref{fig:cont_phase} (red) reveals that there is far more
information about the Jeans scale in quasar pair spectra than can be
measured with the cross-power. The cross-power produces constraints
which are effectively a hybrid between the horizontal Jeans scale
contours for the phase angle distribution and the diagonal banana
shaped contours produced by the longitudinal power, which reflects the
degeneracy between Jeans smoothing and thermal broadening.  This
quantitatively confirms our argument in \S~\ref{cps_vs_phase}, that
the cross-power is a product of moduli, containing information about
the 1D power, and the cosine of the phase, which depends on the 3D
power.

\begin{figure}
  \centering
  \centerline{\epsfig{file=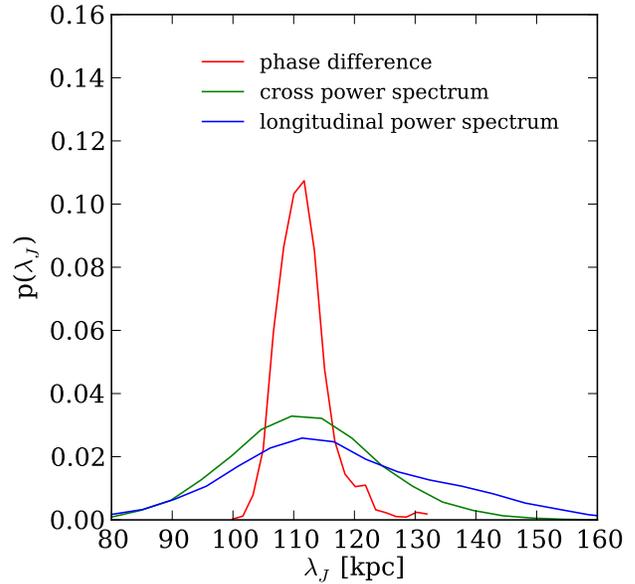,
      width=0.5\textwidth}}
  \vskip -0.1in

  \caption{ \label{fig:marg_l} Estimated accuracy on the measurement
    of $\lambda_J$, obtained marginalizing over $T_0$ and $\gamma$ the 
    posterior distribution from the MCMC analysis.  The phase
    difference statistic (red) sets tighter constraints than the cross
    power (blue) and the longitudinal power (black), which are
    affected by parameter degeneracies. In this case we do not 
    account for the effect of noise and limited resolution, and we 
    find a relative precision of 3.9\% for $\lambda_J$.}
\end{figure}

The results of this section indicate that among the statistics that we
have considered, the phase angle PDF is the most powerful for
constraining the IGM pressure smoothing, because it is more sensitive
to the Jeans scale and results in constraints that are free of
degeneracies with other thermal parameters.  
We demonstrate this
explicitly in Figure~\ref{fig:marg_l}, where we show the fully
marginalized posterior distribution (see eqn.~\ref{fulllik}) of the
Jeans scale for each the statistics we have considered. The
probability distributions quantify the visual impression from the
contours in Figures~\ref{fig:cont_phase} and \ref{fig:cont_cps}, and
clearly indicate that the phase angle PDF is the most sensitive. 
The relative error on the Jeans scale
$\sigma_{\lambda}\slash \lambda_J = 3.9\%$, which is a remarkable
precision when compared to the typical precision $\sim 30\%$ of
measurements of $T_0$ and $\gamma$ in the published literature
\citep[see e.g. Figure 30 in][for a recent compilation]{Lidz09},
especially when one considers that only 20 quasar pair spectra are
required to achieve this accuracy. 

We close this section with a caveat to our statements that our Jeans
scale constraints are free of degeneracies with other thermal
parameters. The phase angle PDF is \emph{explicitly} nearly
independent of the temperature-density relation because 1) the
non-linear FGPA transformation is only weakly dependent on temperature
and 2) the phase angle PDF is invariant to the thermal broadening
convolution (see \S~\ref{sec:flux}). However, in our idealized dark-matter only simulations, the
Jeans scale is taken to be completely independent of $T_0$ and
$\gamma$; whereas, in reality all three parameters are correlated by
the underlying thermal history of the Universe. In this regard, the
Jeans scale may \emph{implicitly} depend on the $T_0$ and
$\gamma$ at the redshift of the sample, as well as with their values
at earlier times. We argued that because the thermal history is not
known, taking the Jeans scale to be free parameter is reasonable.
However, the validity of this assumption and the implicit dependence of the
Jeans scale on other thermal parameters is clearly something that
should be explored in the future with hydrodynamical simulations. 

\subsection{The Impact of Noise and Finite Spectral Resolution}
\label{sec:noise}

Up until this point we have assumed perfect data with infinite
signal-to-noise ratio and resolution. This is unrealistic, especially
considering, as discussed in \S~\ref{sec:mock}, that that close quasar
pairs are faint, and typically cannot be observed at echelle resolution
or very high signal-to-noise ratio $\gtrsim 20$, even with 8m class
telescopes.  In this section we explore the impact of noise and finite
resolution on the precision with which we can measure the Jeans scale.

We consider the exact same sample of 20 mock quasar spectra, but now
assume that they are observed with spectral resolution corresponding
to FWHM $=30\,\kms$, and two different signal-to-noise ratios of ${\rm S\slash N}
\simeq 5$ and ${\rm S\slash N} \simeq 10$ \emph{per pixel}.  These
values are consistent with what could be achieved using an echellette
spectrometer on an 8m class telescope. To create mock observed spectra with these
properties, we first smooth our simulated spectra with a Gaussian kernel to model
the limited spectral resolution, and interpolate these smoothed spectra onto a coarser
spectral grid which has $10\,\kms$ pixels, consistent with the spectral pixel scale
of typical echellette spectrometers. We then add Gaussian white noise to each pixel with 
variance $\sigma^2_{\rm N}$ determined by the relation ${\rm S\slash N}=\bar{F}/\sigma_{\rm N}$, 
where $\bar{F}$ is the mean transmitted flux. This then gives an average signal-to-noise ratio
equal to the desired value. 

As we already discussed in \S~\ref{sec:flux} in the context of thermal
broadening, phase angles are invariant under a convolution with a
symmetric Gaussian kernel.  Thus we do not expect spectral resolution
to significantly influence our results, provided that we restrict
attention to modes which are marginally resolved, such that we can
measure their phases.  Indeed, the cutoff in the flux power spectrum
induced by spectral resolution is $k_{res}=1/\sigma_{\rm res}\approx
2.358/{\rm FWHM} = 0.08\,\skm$, is comparable to the maximum wavenumber we
consider $k=0.1\,\skm$, and hence we satisfy this criteria. Note
further that this invariance to a symmetric spectral convolution
implies that we do not need to be able to precisely model the
resolution, provided that it has a nearly symmetric shape and does
not vary dramatically across the spectrum.  This is another significant
advantage of the phase angle approach, since the resolution of a
spectrometer often depends on the variable seeing, and can be
challenging to accurately calibrate.

\begin{figure}
  \centering
  \centerline{\epsfig{file=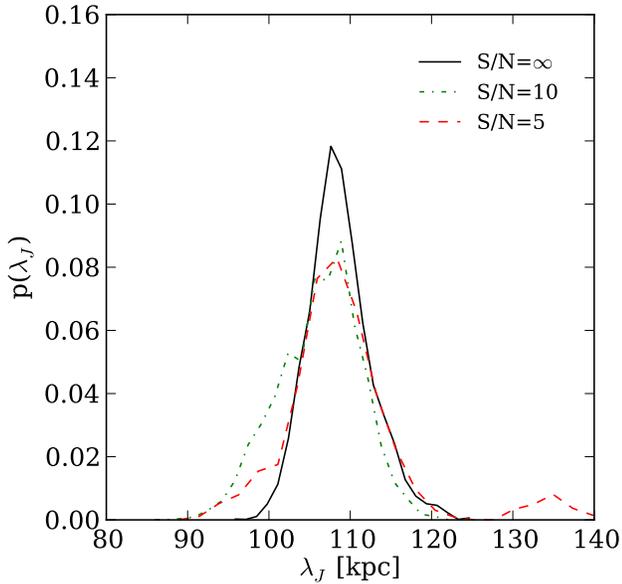,
      width=0.5\textwidth}}
  \vskip -0.1in
  
  \caption{\label{fig:marg_l_noise} The effect of noise and 
resolution in the measurement of $\lambda_J$. The plots shows
the posterior distribution of the Jeans scale, marginalized 
over $T_0$ and $\gamma$. Each line represent a different degree 
of noise, assuming a resolution of FWHM=30 km/s. We selected a 
different subsample of the simulation as our mock dataset which
has a precision of 3.6\% for S/N=$\infty$ (black solid), 4.8\% for 
S/N=$10$ (green dot-dashed) and 7.2\% for S/N=$5$ (red dashed). }
\end{figure}

Although our results are thus likely to be very independent of
resolution, noise introduces fluctuations which are uncorrelated
between the two sightlines, and this will tend to reduce the coherence
of the flux that the phase angle PDF quantifies. Noise will  
thus modify the shape of the phase angle PDF away from the intrinsic shape shown in
Figure~\ref{fig:fluxphase}. In order to deal with noise and its
confluence with spectral resolution, we adopt a forward-modeling
approach.  Specifically, for each thermal model we smooth all 10,000
IGM skewers to finite resolution, interpolate onto coarser spectral
grids, and add noise consistent with our desired signal-to-noise
ratio. We then fit the resulting distribution of phase angles to the
wrapped-Cauchy distribution, determining the value of the
concentration parameter $\zeta(k,r_{\perp})$, at each $k$ and
$r_{\perp}$ as we did before. We again emulate the function
$\zeta(k,r_\perp|T_0,\gamma, \lambda_J)$ using the same thermal
parameter grid, but now with noise and spectral resolution included,
enabling fast evaluations of the likelihood in
eqn.~(\ref{diaglik}). Thermal parameter constraints then follow from MCMC 
exploration of this new likelihood, for which the impact of noise and resolution
on the phase angle PDF have been fully taken into account. 

In Figure~\ref{fig:marg_l_noise} we show the impact of noise on the
fully marginalized constraints on the Jeans scale from the phase angle
PDF. The solid curve represents the posterior distribution for a mock
dataset with infinite resolution and signal-to-noise ratio, which is
identical to the red curve in Figure~\ref{fig:marg_l}. The dotted and
dashed curves illustrate the impact of ${\rm S\slash N}=10$ and ${\rm
  S\slash N}=5$, respectively. Note that the slight shift in the modes
of these distributions from the fiducial value are expected, and
should not be interpreted as a bias. Different noise realizations
generate scatter in the phase angles just like the intrinsic noise
from large-scale structure. The inferred Jeans scale for any given
mock dataset or noise realization will not be exactly equal to the
true value, but should rather be distributed about it with a scatter
given by the width of the resulting posterior distributions. The
relative shifts in the mode of the posterior PDFs are well within
$1\sigma$ of the fiducial value, and are thus consistent with our
expectations. 

The upshot of Figure~\ref{fig:marg_l_noise} is that noise and limited
spectral resolution do not have a significant impact on our ability to
measure the Jeans scale. For a signal-to-noise ratio of ${\rm S\slash
  N}=10$ per pixel we find that the relative precision with which we
can measure the Jeans scale is $\sigma_{\lambda}\slash \lambda_J
=4.8\%$, which is only a slight degradation from the precision
achievable from the same dataset at infinite signal-to-noise ratio
and resolution $\sigma_{\lambda}\slash \lambda_J=3.9\%$.  The
small impact of noise on the Jeans scale precision is not
surprising. For the $10\,\kms$ spectral pixels that we simulate, the
standard deviation of the normalized Ly$\alpha$ forest flux per pixel is
$\sqrt{\langle \delta F^2\rangle} \simeq 32\%$, 
whereas for ${\rm S\slash
  N}=10$ our Gaussian noise fluctuations are at a significantly
smaller $\simeq 10\%$ level. Heuristically, these two `noise' sources
add in quadrature, and thus the primary source of `noise' in measuring
the phase angle PDF results from the Ly$\alpha$ forest
itself, rather than from noise in the data. For a lower
signal-to-noise ratio of ${\rm S\slash N}=5$ per pixel, the precision is further
degraded to $\sigma_{\lambda}\slash \lambda_J=7.2\%$, which reflects the
fact that noise fluctuations are becoming more comparable to the intrinsic
Ly$\alpha$ forest fluctuations. 

These numbers on the scaling of our precision with signal-to-noise
ratio ${\rm S\slash N}$ provide intuition about the optimal observing
strategy. For a given sample of pairs, it will require four times more
exposure time to increase the signal-to-noise ratio from ${\rm S\slash
  N}\simeq5$ to ${\rm S\slash N}\simeq10$, whereas the same telescope time allocation
could be used to increase the sample size by a factor of four at the
same signal-to-noise ratio (assuming sufficient close pair sightlines
exist). For the latter case of an enlarged sample, the precision will
scale roughly as $\propto \sqrt{N_{\rm pairs}}$, implying a
$\sigma_{\lambda}\slash \lambda_J=3.6\%$ for a sample of 80 pairs
observed at ${\rm S\slash N}=5$. This can be compared to
$\sigma_{\lambda}\slash \lambda_J =4.8\%$ for 20 pairs observed at
${\rm S\slash N}\simeq10$. There is thus a marginal
gain in working at lower ${\rm S\slash N}\simeq 5$ and observing a larger
pair sample, although we have not considered various systematic errors
which could impact our measurement. However, higher signal-to-noise spectra are
usually  preferable for the purposes of mitigating systematics, and
hence one would probably opt for higher signal-to-noise ratio, a smaller pair sample, and 
tolerate slightly higher statistical errors.

\subsection{Systematic Errors}

\begin{figure}
  \centering
  \centerline{\epsfig{file=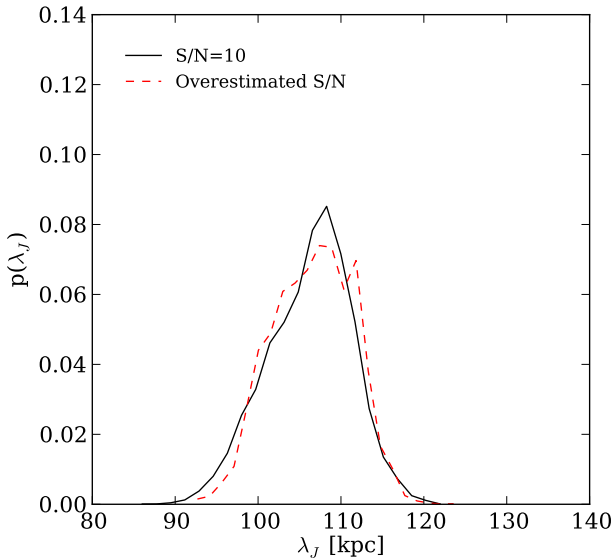,
      width=0.5\textwidth}}
  \vskip -0.1in
  
  \caption{\label{fig:wrong_noise} The effect of overestimating the signal-to-noise 
  ratio by a 20\% factor (red, dashed line) when the real value is S/N$=10$: we do not 
  find any significant bias on the measured value of the Jeans scale.}
\end{figure}

We now briefly discuss the systematic errors which could impact a
measurement of the Jeans scale. First, consider the impact of errors
in the continuum normalization. Because the phase angle is a ratio of
Fourier modes of the normalized flux eqn.~(\ref{eqn:phase}), it is
completely insensitive to the continuum normalization of $\delta F$,
provided that the continuum is not adding significant power on the
scale of wavelength of the $k$-mode considered. In the previous section, we argued
that finite spectral resolution does not have a significant impact the phase angle PDF, because 
phase angles are invariant under convolutions with symmetric kernels. We do take resolution
into account in our forward-modeling of the phase angle PDF, but precise knowledge of the
spectral resolution or the line spread function is not required,
since the line spread function will surely be symmetric when averaged
over several exposures, thus leaving the phase angles invariant. The
only requirement is that we restrict attention to modes less than the
resolution cutoff $k \lesssim k_{\rm res}$ whose amplitudes are not
significantly attenuated, such that we can actually measure their
phase angles.

Noise does modify the phase angle PDF, but our forward-modeling
approach takes this fully into account provided the noise estimates
are correct. One potential systematic is uncertainty in the noise
model. The typical situation is that the standard-deviation of a spectrum
reported by a data reduction pipeline is underestimated at the $\sim
10-20\%$ level (${\rm S\slash N}$ overestimated), because of systematic
errors related to the instrument and data reduction \citep[see
  e.g.][]{msb+06,KGLeeBOSS13}. To address this issue we directly
model the impact of underestimated noise for a case where we think the
${\rm S\slash N}\simeq 10$, but where in reality it is actually $20\%$ lower ${\rm
  S\slash N}\simeq 8$. Specifically, using our same mock dataset we
generate 20 quasar pair spectra with ${\rm S\slash N}\simeq 8$. However,
when forward-modeling the phase angle PDF with the IGM simulations, we
take the signal-to-noise ratio to be the overestimated value of ${\rm
  S\slash N}\simeq 10$. Excess noise above our expectation would tend to
reduce the coherence in the spectra (less peaked phase angle PDF)
mimicking the effect of a smaller Jeans scale. We thus expect a bias
in the Jeans scale to result from the underestimated
noise. Figure~\ref{fig:wrong_noise} compares the posterior
distributions of the Jeans scale for the two cases 
${\rm S\slash N}\simeq 10$ (black curve) and signal-to-noise ratio
overestimated to be ${\rm S\slash N}\simeq 10$ but actually equal to ${\rm
  S\slash N}\simeq 8$ (red curve). We see that $\simeq 20\%$ level
uncertainties in the noise lead to a negligible bias in the Jeans
scale. 

The only remaining systematic that could impact the Jeans scale
measurement is metal-line absorption within the forest. Metal
absorbers cluster differently from the IGM, and it is well known that
metals add high-$k$ power to the Ly$\alpha$ forest power spectrum 
because the gas traced by metal lines tends to be colder than 
\ion{H}{1} in the IGM \citep{McDonald2000,Croft2002,Kim04,Lidz09}. As this
metal absorption is not present in our IGM simulations, it can lead to
discrepancies between model phase angle PDFs and the actual data,
resulting in a biased measurement. This is very
unlikely to be a significant effect. We restrict attention to
large scale modes with $k < 0.1\,\skm$, both because this is comparable
to our expected spectral resolution cutoff, and because below these
wavenumbers metal line absorption results in negligible contamination
of the longitudinal power \citep{McDonald2000,Croft2002,Kim04,Lidz09}. Since the metal absorbers
have a negligible effect on the \emph{moduli} of these large scale modes, we also 
expect them to negligibly change their phase angles. 

We thus conclude that the phase angle PDF is highly insensitive to the
systematics that typically plague Ly$\alpha$ forest measurements, such
as continuum fitting errors, lack of knowledge of spectral resolution,
poorly calibrated noise, and metal line absorption.

\subsection{Is Our Likelihood Estimator Unbiased?}\label{sec:bias}

Finally, we determine whether our procedure for measuring the Jeans
scale via the phase angle likelihood (eqn.~\ref{diaglik}) outlined at
the end of \S~\ref{sec:emulator}, produces unbiased estimates. To
quantify any bias in our Jeans scale estimator we follow a Monte Carlo
approach, and generate 400 distinct quasar pair samples by randomly
drawing 20 quasar pair spectra (allowing for repetition) from our
ensemble of 10,000 skewers. Note that the distribution of transverse
separations is approximately the same for all of these realizations,
since we only simulate 30 discrete separations, and the full sample of
20 overlapping pair spectra requires 200 pairs of skewers, which are
randomly selected from among the 30 available pair separations. We
MCMC sample the likelihood in eqn.~(\ref{diaglik}) for each
realization, and thus generate the full marginalized posterior
distribution (eqn.~\ref{fulllik}; red curve in
Figure~\ref{fig:marg_l}).  The `measured' value of the Jeans scale for
each realization is taken to the be the mean of the posterior
distribution. We conducted this procedure for the case of finite
spectral resolution (FWHM $=30\,\kms$) and signal-to-noise ratio ${\rm
  S\slash N}\simeq 5$, where our forward-modeling procedure described
in \S~\ref{sec:noise} is used to model the impact of resolution and
noise on the phase angle PDF.

\begin{figure}
  \centering
  \centerline{\epsfig{file=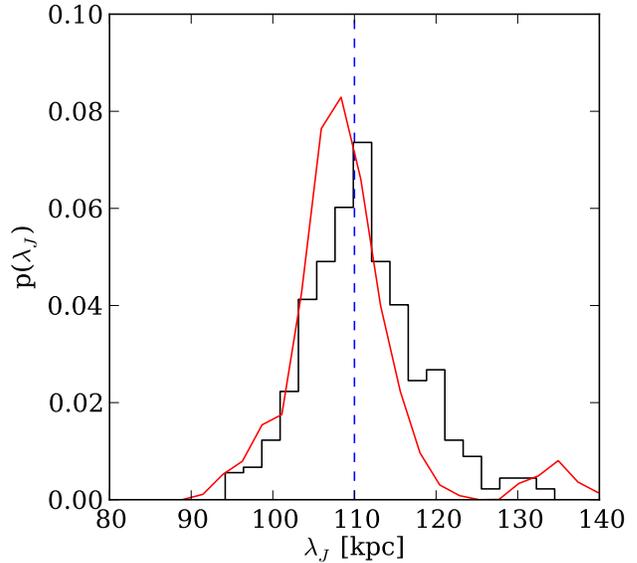,
      width=0.5\textwidth}}
  \vskip -0.1in

  \caption{  \label{fig:bias} Probability distribution of the measured value of $\lambda_J$ 
  for 400 different mock datasets drawn from the fiducial simulation. 
  This plot confirms that our method is not biased, since 
  the distributions is be centered at the true value,
  marked with a vertical dashed line. This test is performed assuming 
  S/N$=5$. The red line is the posterior distribution deduced from our MCMC sampling of the
  phase angle PDF likelihood for one of these  400 mock dataset realizations. Its
  similarity in shape to the distribution of mock measurements illustrates that our
  MCMC simulations provide reliable error estimates.}
\end{figure}

The distribution of Jeans scale measurements resulting from this Monte
Carlo simulation is shown in Figure~\ref{fig:bias}. We find that the distribution of
'measurements' is well centered on the true value of $\lambda_J=110$
kpc, and the mean value of this distribution is $\lambda_J=111.1$ kpc,
which differs from the true value by only $1\%$, confirming that our
procedure is unbiased to a very high level of precision. The relative
error of our measurements from this Monte Carlo simulation is
$\sigma_{\lambda_J}\slash \lambda_J=6.3\%$, which is consistent with
the value of $\sigma_{\lambda_J}\slash \lambda_J=7.2\%$, which we
deduced in \S~\ref{sec:accuracy} from an MCMC sampling of the
likelihood for a single mock dataset. This confirms that the posterior
distributions derived from our MCMC do indeed provide an accurate
representation of the errors on the Jeans scale and other thermal
parameters. However, we note that there is some small variation in the
value of $\sigma_{\lambda_J}\slash \lambda_J$ inferred from the posterior distributions 
for different mock data realizations, as expected. Given that we only generated 400 samples,
the error on our determination of the mean of the distribution in
Figure~\ref{fig:bias} is $\simeq \sigma_{\lambda_J}\slash
\lambda_J\slash \sqrt{400} = 0.3\%$, and thus our slight bias of $1\%$
constitutes a $\sim 3\sigma$ fluctuation. We suspect that this is too
large to be a statistical fluke, and speculate that a tiny
amount of bias could be resulting from interpolation errors in our
emulation of the IGM. It is also possible that choosing an alternative
statistic of the posterior distribution as our `measurement' instead
of the mean, for example the mode or median, could also further reduce
the bias. But we do not consider this issue further, since the bias is so small
compared to our expected precision.

We conclude that our phase angle PDF likelihood procedure for estimating
the Jeans scale has a negligible $\simeq 1\%$ bias. We would need to
analyze a sample of $\simeq 500-1000$ quasar pair spectra for this
bias to be comparable to the error on the Jeans scale. Furthermore, it
is likely that we could, if necessary, reduce this bias even further by
either reducing the interpolation error in our emulator or by applying
a different statistic to our posterior distribution to determine the
measured value.

\section{Discussion and Summary}\label{summary}
Spectra of correlated Ly$\alpha$ forest absorption in close quasar
pair sightlines represent a unique opportunity to improve our
understanding of the physics governing the IGM.  In this paper we have
shown that the degree of coherence of Ly$\alpha$ absorption in quasar
pair spectra is sensitive to the Jeans filtering scale, provided the
pair separation is small enough to resolve it. Although the Jeans
scale has never been measured, it has fundamental cosmological
implications: it provides a thermal record of heat injected by
ultraviolet photons during cosmic reionization events, determines the
clumpiness of the IGM, a critical ingredient in reionization models,
and sets the minimum mass of galaxies to gravitationally collapse from
the IGM.

We introduce a novel technique to directly measure the Jeans scale
from quasar pair spectra based on the probability distribution
function (PDF) of phase angle differences of homologous longitudinal
Fourier modes in close quasar pair spectra. To study the efficacy of
this new method, we combined a semi-analytical model of the
\mlya\ forest with a dark matter only simulation, to generate a grid
of 500 thermal models, where the temperature at mean density $T_0$,
slope of the temperature-density relation $\gamma$, and the Jeans
scale $\lambda_J$ were varied.  A Bayesian formalism is introduced
based on the phase angle PDF, and MCMC techniques are used to conduct
a full parameter study, allowing us to characterize the precision of a
Jeans scale measurement, explore degeneracies with the other thermal
parameters, and compare parameter constraints with those obtained from
other statistics such as the longitudinal power and the cross-power
spectrum.

The primary conclusions of this study are: 

\begin{itemize}
\item The longitudinal power is highly degenerate with
  respect to the thermal parameters $T_0$, $\gamma$ and $\lambda_J$,
  which arises because thermal broadening smooths the IGM along the
  line-of-sight (1D) at a comparable scale as the Jeans pressure
  smoothing (3D).  It is extremely challenging to disentangle this
  confluence of 1D and 3D smoothing with longitudinal observations
  alone. Similar analogous degeneracies are likely to exist in other
  previously considered statistics sensitive to small-scale power such
  as the wavelet decomposition, the curvature, the $b$-parameter
  distribution, and the flux PDF.  Hence it may be necessary to
  reassess the reliability and statistical significance of previous
  measurements of $T_0$ and $\gamma$.

 \item The cross-power measured from close quasar pairs is
   sensitive to the 3D Jeans smoothing, and can break 
   degeneracies with the unknown Jeans scale. However, it is not the
   optimal statistic, because it mixes 1D information in the moduli of
   longitudinal Fourier modes, with the 3D information encoded in their phase
   differences. We show that by focusing on the phase differences 
   alone, via the full PDF of phase angles, one is much more sensitive to 3D power and 
   the Jeans smoothing.

 \item Based on a simple heuristic geometric argument, we derived an
   analytical form for the phase angle PDF. A single parameter family
   of wrapped-Cauchy distributions provides a good fit to the phase
   differences in our simulated spectra for any $k$, $r_{\perp}$, the
   full range of $T_0$,$\gamma$ and $\lambda_J$.

\item Our phase angle PDFs indicate that phase differences between
  large-scale longitudinal modes with small wavenumbers $k \ll
  1/\lambda_J$, are nevertheless very sensitive to the Jeans scale. We
  present a simple analytical argument showing that this sensitivity
  results from the geometry of observing a 3D field along 1D
  skewers: low-$k$ cross-power across correlated 1D skewers is
  actually dominated by high-$k$ 3D modes up to a scale set by the
  pair separation $k_\perp \sim 1\slash r_\perp$.

\item The phase angle PDF is essentially independent of the
  temperature-density relation parameters $T_0$ and $\gamma$. This
  results because 1) the non-linear FGPA transformation is only weakly
  dependent on temperature 2) phase angles of longitudinal modes are invariant to the
  symmetric thermal broadening convolution.

\item Our full Bayesian MCMC parameter analysis indicates that a
  realistic sample of only 20 close quasar pair spectra observed at
  modest signal-to-noise ratio ${\rm S\slash N}\simeq 10$, can pinpoint the
  Jeans scale to $\simeq 5\%$ precision, fully independent of the
  amplitude $T_0$ and slope $\gamma$ of the temperature-density
  relation.  The freedom from degeneracies with $T_0$ and $\gamma$ is a direct consequence
  of the near independence of the phase angle PDF of these parameters.

\item Our new estimator for the Jeans scale is unbiased and
  insensitive to a battery of systematics that typically plague
  Ly$\alpha$ forest measurements, such as continuum fitting errors,
  imprecise knowledge of the noise level and/or spectral resolution,
  and metal-line absorption.

\end{itemize}

In order for the parameter study presented here, with a large grid
(500) of thermal models, to be computationally feasible, we had to
rely on a simplified model of the IGM, based on a dark-matter only
simulation and simple thermal scaling relations. In particular, the
impact of Jeans pressure smoothing on the distribution of baryons is
approximated by smoothing the dark-matter particle distribution with a
Gaussian-like kernel, and we allowed the three thermal parameters
$T_0$, $\gamma$, and $\lambda_J$ to vary completely independently.
Although the Gaussian filtering approximation is valid in linear
theory \citep{GnedinBaker2003}, the Jeans scale is highly nonlinear at
$z\simeq 3$, hence a precise description of how pressure smoothing
alters the 3D power spectrum of the baryons requires full
hydrodynamical simulations. Furthermore, the three thermal parameters
we consider are clearly implicitly correlated by the underlying
thermal history of the Universe. Indeed, a full treatment of the
impact of impulsive reionization heating on the thermal evolution of
the IGM and the concomitant hydrodynamic response of the baryons,
probably requires coupled radiative transfer hydrodynamical
simulations.  

Our approximate IGM model is thus justified by the
complexity and computational cost of fully modeling the Jeans smoothing problem, and
despite its simplicity, it provides a good fit to current measurements
of the longitudinal power (see Figure~\ref{fig:power_spectra}). Most
importantly, our simple model allowed us to develop valuable physical
intuition about how 3D pressure smoothing of baryons is manifest in
Ly$\alpha$ forest spectra of close quasar pairs. Based on this
intuition, we devised a powerful new method which isolates this
small-scale 3D information.  By combining this new technique with
existing close quasar spectra, we will make the first direct
measurement of the Jeans scale of the IGM. Given that precise $\simeq 5\%$
constraints on the Jeans scale will soon be available, the time is
now ripe to use hydrodynamical and radiative transfer simulations to
improve our understanding of how reionization heating altered the
small-scale structure of baryons in the IGM.

\acknowledgments 

We thank P. McDonald and U. Seljak for first suggesting to JFH that
close quasar pairs could be used to measure the Jeans scale.  We also
thank the members of the ENIGMA
group\footnote{http://www.mpia-hd.mpg.de/ENIGMA/} at the Max Planck
Institute for Astronomy (MPIA) for reading an early version of the
manuscript and for helpful discussions. JFH acknowledges generous
support from the Alexander von Humboldt foundation in the context of
the Sofja Kovalevskaja Award. The Humboldt foundation is funded by the
German Federal Ministry for Education and Research.

\bibliographystyle{../Bibli/apj}
\bibliography{../Jeans/jeans}

\appendix
\section{Resolving the Jeans Scale with Dark-Matter Simulations}
\label{sec:appendixa}

The \mlya\ forest probes the structure of the very low density
regions of the IGM, setting strict requirements on the
resolution of our dark-matter only simulation.  In particular, because
our simulation is discrete in mass, each dark-matter particle
represents a fixed amount of gas distributed according to the
gravitational softening length and the size of the smoothing kernel
that we use to represent Jeans smoothing (eqn.~\ref{eqn:kernel}).  At
very low densities, it is possible that a very large region is
described by a single particle, and that most of this void region is
left empty. This undesirable situation occurs when the mean
inter-particle separation $\Delta l = L_{\rm box} \slash N_{\rm p}^{1\slash
  3}$, which defines the typical size of regions occupied by each
particle, is much larger than the Jeans scale $\lambda_J$, which is
the minimum scale we want to resolve.  Under such circumstances the
density profile of skewers through our simulation cube will have many
pixels which are nearly empty, because they have few or no neighboring
particles. This insufficient sampling of the volume due to large
mean inter-particle separation will then manifest itself through the
appearance of artifacts in the volume-weighted probability
distribution function (PDF) of the density.  On the other hand, if the
inter-particle separation is sufficiently small, the density field
will be sufficiently sampled, and further decreasing the
inter-particle separation will not alter the density PDF. Therefore we
can define our resolution criteria for the mean inter-particle separation
to be smaller than some multiple of the Jeans scale $\Delta l < \alpha
\lambda_J$, where the exact value of this coefficient $\alpha$ is
determined by checking that convergence is achieved in the density
PDF.

\begin{figure}
  \centering
  \centerline{\epsfig{file=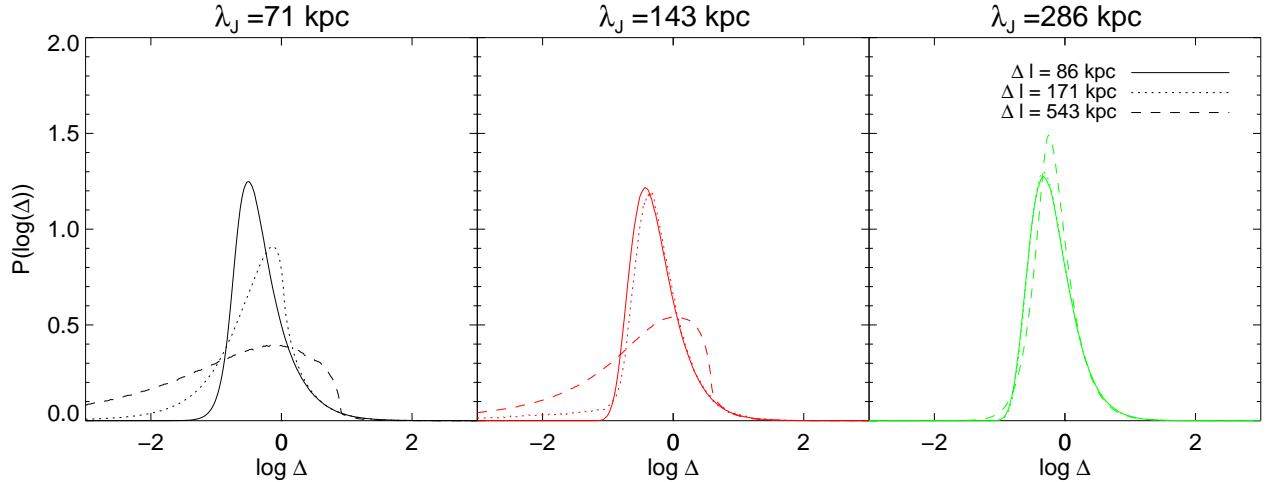,
      width=\textwidth}}
  \vskip -0.1in
  
  \caption{\label{fig:conv_test} The probability distributions of the relative baryonic 
  density $\Delta=\rho/\bar{\rho}$ in our simulations. Each panel represent a different 
  filtering scale $\lambda_J$, which was used to smooth the dark matter density for the same three 
  simulations, which have different mean inter-particle separations $\Delta l$. When $\Delta l$ is too 
  large relative to $\lambda_J$ the IGM density is poorly resolved at low densities, and the PDF is not 
  converged. Empirically, we find that a safe criterion for convergence is $\Delta l < \lambda_J$, 
  which allow us to resolve Jeans scales down to 50 kpc with our $L_{\rm box} = 50\,h^{-1}\,{\rm Mpc}$ and 
  $N_{\rm p}=1500^3$ simulation.}
\end{figure}

We estimate $\alpha$ by plotting the PDF of log$(\Delta)$ from our IGM
skewers for a set of simulations with varying mean inter-particle
separation, where $\Delta = \rho/ \bar{\rho} = 1 + \delta$ is the
density in units of the mean. The employed simulations have mean
inter-particle separations $\Delta l = \{ 86 , 171 , 653\}$ kpc, corresponding to box sizes 
$L_{\rm box}=\{100 , 250 , 720\}$ Mpc$/h$ with $N_{\rm p} = \{ 1500^3, 2048^3 , 1800^3\}$ particles, 
respectively. In Figure~\ref{fig:conv_test} we check for convergence using
three different values of $\lambda_J$. The results indicate that a safe
criterion for resolving the jeans scale is $\Delta l < \lambda_J$ or $\alpha \simeq 1$. 
The simulation employed in this work has $L_{\rm box} = 50\,h^{-1}\,{\rm Mpc}$ and $N_{\rm p}=1500^3$ 
particles, or a mean inter-particle separation of $\Delta l = 48\,{\rm kpc}$. This simulation 
thus allows us to study pressure smoothing down to a Jeans scale as small as $\simeq 50\,{\rm kpc}$. 
Note however that the results of this paper rely on our estimation of the Jeans scale from various 
Ly$\alpha$ forest statistics around the fiducial value of $\lambda_J=110\,{\rm kpc}$, so we are confident
that the Jeans scale is resolved in our simulations and that our results are not impacted by resolution 
effects. 

\section{Determining the Concentration Parameter $\zeta$ of the Wrapped-Cauchy Distribution}\label{rho_determin}
For a given sample of phases $\{\theta\}$ we employ a maximum-likelihood algorithm
to determine the best-fit concentration parameter $\zeta$, which uniquely specifies a
wrapped-Cauchy distribution. This procedure is described in detail in
\citet{Jammalamadaka}.  Briefly, we first reparametrize the wrapped-Cauchy distribution (eqn.~\ref{WCD}) by
writing $\nu=2\zeta/(1+\zeta^2)$, which gives
\begin{equation}
 P(\theta)\propto\frac{1}{1-\nu \cos(\theta)} \equiv w(\theta | \nu).
\end{equation}
Following the standard recipe of maximizing the logarithm of the 
likelihood with respect to the desired parameter, 
we sum the logarithms of the probability of all angles 
and impose the condition that its derivative with respect to $\nu$
is zero, resulting in the equation
\begin{equation}
\sum_{i=1}^{n} w(\theta_i|\nu)[\cos(\theta_i)-\nu]=0, 
\end{equation}
which can be solved iteratively. The concentration parameter is then
easily determined by inverting the above relation to get
$\zeta=(1-\sqrt{1-\nu^2})/ \nu$.  This procedure is repeated for
each distinct population of phases, parametrized by transverse
separation $r_{\perp}$ and $k$-mode, $\theta(r_{\perp},k)$, and for
each model in the thermal parameter grid
$(T_0,\gamma,\lambda_J)$ that we consider.

\end{document}